\documentclass[a4paper,10pt]{amsart}
\usepackage{graphicx}
\usepackage{subfigure}

\usepackage{amssymb, amsmath}
\usepackage{tikz}
\usepackage{young}
\usepackage{youngtab}
\usepackage{array}

\usepackage[margin=2.5 cm]{geometry}

\newcommand{\modulus}{\mbox{ \textsc{mod} }}
\newcommand{\tmodulus}{\tilde{\mbox{ \textsc{mod} }}}
\newtheorem{theorem}{Theorem}

\def\RADIUS{1cm}
\def\VA{(0:\RADIUS)}%
\def\VB{(30:\RADIUS)}%
\def\VC{(60:\RADIUS)}%
\def\VD{(90:\RADIUS)}%
\def\VE{(120:\RADIUS)}%
\def\VF{(150:\RADIUS)}%
\def\VG{(180:\RADIUS)}%
\def\VH{(210:\RADIUS)}%
\def\VI{(240:\RADIUS)}%
\def\VJ{(270:\RADIUS)}%
\def\VK{(300:\RADIUS)}%
\def\VL{(330:\RADIUS)}%

\def\dodecagon{
\draw \VA circle (1pt) node[right]{\tiny $1$};%
\draw \VB circle (1pt) node[above right]{\tiny $8$};%
\draw \VC circle (1pt) node[above right]{\tiny $3$};%
\draw \VD circle (1pt) node[above]{\tiny $10$};%
\draw \VE circle (1pt) node[above left]{\tiny $5$};%
\draw \VF circle (1pt) node[above left]{\tiny $12$};%
\draw \VG circle (1pt) node[left]{\tiny $7$};%
\draw \VH circle (1pt) node[below left]{\tiny $2$};%
\draw \VI circle (1pt) node[below left]{\tiny $9$};%
\draw \VJ circle (1pt) node[below]{\tiny $4$};%
\draw \VK circle (1pt) node[below right]{\tiny $11$};%
\draw \VL circle (1pt) node[below right]{\tiny $6$};%
}

\newcommand{\twelve}{\hbox{$12$}}
\newcommand{\eleven}{\hbox{$11$}}
\newcommand{\ten}{\hbox{$10$}}

\begin{document}

\title[Understanding complex dynamics by means of an associated Riemann surface]{Understanding complex dynamics by means of an associated Riemann surface}%

\author[D. Gomez-Ullate]{David G\'omez-Ullate}%
\address{Departamento de F\'isica Te\'orica II, Universidad Complutense, Madrid, Spain.}%
\email[D. Gomez-Ullate]{david.gomez-ullate@fis.ucm.es}

\author[P. M. Santini]{Paolo Maria Santini}
\address{Dipartimento di Fisica, Universit\`{a} di Roma ``La Sapienza'', Roma, Italy.}%
\noindent
\address{Istituto Nazionale di Fisica Nucleare, Sezione di Roma, Italy.}%
\email[P. M. Santini]{paolo.santini@roma1.infn.it}

\author[M. Sommacal]{Matteo Sommacal}
\address{Institut des Hautes Etudes Scientifiques, Bures-sur-Yvette, France.}%
\email[M. Sommacal]{sommacal@ihes.fr}

\author[F. Calogero]{Francesco Calogero}
\address{Istituto Nazionale di Fisica Nucleare, Sezione di Roma, Italy.}%
\email[F. Calogero]{francesco.calogero@roma1.infn.it}

\date{\today}




\begin{abstract}

We provide an example of how the complex dynamics of a recently
introduced model can be understood via a detailed analysis of its
associated Riemann surface. Thanks to this geometric description an
explicit formula for the period of the orbits can be derived, which
is shown to depend on the initial data and the continued fraction
expansion of a simple ratio of the coupling constants of the
problem. For rational values of this ratio and
generic values of the initial data, all orbits are periodic and the system is isochronous.
For irrational values of the ratio, there exist periodic and quasi-periodic orbits for different initial data.
Moreover, the  dependence of the period on the initial data shows a rich behavior and initial data can always be found such the period is arbitrarily high.
\end{abstract}

\maketitle

\section{Introduction}\label{Sec:Introduction}

Although the evolution of physical systems takes place over real
time and can be described by real variables, the interest of
extending the study to complex values has long been noticed. For
instance, the solutions of a nonlinear system of ODEs representing a
physical phenomenon might have singularities off the real time axis
whose position depends on the initial data. Even though the time
variable is followed along the real axis, the position of these
complex singularities provide information on the evolution of the
system and its dependence on the choice of initial data.

Historically, one of the first successes of this paradigm is the
collection of techniques now known as Painlev\'e analysis,
originally introduced by Painlev\'e and Ko\-wa\-levs\-ka\-ya
\cite{painleve,kow2}. In essence, they consider an \emph{ansatz} of
the local behaviour of a solution near a singularity in terms of a
Laurent series, introducing it in the equations and determining the
leading orders and resonances (terms in the expansion at which
arbitrary constants appear). Painlev\'e analysis has been extended
to test for the presence of algebraic branching (weak Painlev\'e
property \cite{RGB89}). These analytic techniques (which have been
algorithmized and are now available in computer packages
\cite{Hereman}) constitute a useful tool in the investigation of
integrability: in many nonlinear systems where no solution in closed
form is known, Painlev\'e analysis provides information on the type
of branching featured by the general solution or by special classes
of solutions. It has also proved useful to identify special values
of the parameters for which generally chaotic systems such as
H\'enon-Heiles or Lorenz are integrable \cite{BSV,CTW82}. The idea that obstruction to integrability is encoded in the branching
properties of solutions and, more precisely, that dense branching is
responsible for non integrability, preventing from extracting any useful
information on existing first integrals, was first put forward by Kruskal
in connection with his poly-Painlev\'e test \cite{KC,kruskal}, a generalization of the
Painlev\'e test allowing one to investigate, in particular, the degree of
multivaluedness of the solutions.

Tabor and his collaborators initiated the study of chaotic systems
from the point of view of the singularity structure od their
solutions \cite{CTW82,WT81, CGTW83,LT88,tabor}. Their local analytic
approach ($\Psi$-series) was complemented by numerical techniques
developed for finding the location of the singularities in complex
time and determining the order of  branching \cite{corliss}. Similar
studies relating singularity structure, chaos and integrability have
been performed by Bountis and his collaborators. They  recognised
that local analysis alone would not be able to characterise global
asymptotic properties of the systems. In a number of papers
\cite{BB,bount} they propose to call \emph{integrable} those cases
in which the Riemann surface has a finite number of sheets, and {\em
non-integrable} if the number is infinite. Using mostly numerical
evidence they conjecture that in the non-integrable cases the
Riemann surfaces are infinitely-sheeted and the projection on the
complex plane of the singularities is dense. Combining analytical
and numerical results for a simple ODE, Bountis and Fokas
\cite{bountis_fokas} have identified chaotic systems with the
property that the singularities of their solutions are dense. More
recently, a research program to investigate classical dynamics
extended to the complex domain is being carried by Bender and his
collaborators, \cite{ABM11,Bender07, Bender10a,Bender10b}. In these
works extensive use is made of numerical explorations to describe
complex and interesting behaviours, which include chaotic systems
\cite{BFHW09},  but the necessity to study the Riemann surfaces is
recognised.

A precise correspondence between dynamical properties of the system
(e.g. chaotic behaviour or sensitive dependence) and specific
geometrical properties of the Riemann surface is still an open
problem, and this paper can be seen as one more step in this
direction.  Our motivation was to introduce a model which is simple
enough that a full description of its Riemann surface can be
performed, yet complicated enough to feature a rich complex
behaviour, whose description would not be feasible by standard
techniques of qualitative theory. Such a model was initially
presented in \cite{CGSS2005}, and in a subsequent paper
\cite{CGSS2008-I} some results were announced without proof since
they required a full description of the Riemann surface, a task
which is carried out in this paper. We stress that achieving the
complete description of the Riemann surface and its consequence for
the determination of the period of the solutions involves the use of
various mathematical techniques ranging from complex analysis to
combinatorics and showing remarkable connections with number theory
(continued fraction expansions).

An important element of the construction of the model is a change of
dependent and independent variables which becomes useful to identify
isochronous systems \cite{C1997,C2001,C2007b}.  Using local analysis
and numerical integration in two many-body systems in the plane
\cite{CS2002,CF02,CFS2003}, it was discovered that outside the
isochrony region there exist periodic solutions with much higher
periods \cite{GS05}, as well as possibly aperiodic solutions, and
the connection among this phenomenology and the analytic structure
of the corresponding solutions as functions of complex time was
illuminated. The model studied in this paper was introduced as a
prototype to understand the complex behaviour with many periodic
orbits observed numerically in the models described above.

Other attempts to study models which are able to produce chaotic
motion and yet lend themselves to a complete description of their
Riemann surfaces are those related to the inversion of hyperelliptic
integrals \cite{FG2006,GS2006}.

This paper is organized as follows: in Section 2 we present the
model, show that it can be reduced to quadratures and how the
general solution can be written in terms of a multi-valued function
that motivates the study of an associated Riemann surface. In
Section 3 we describe the geometric strucure of the Riemann surface,
originally for the algebraic case ($\mu \in \mathbb Q$) . The
connection with graph theory and Ferrer diagrams is illustrated in
Section 4, which  shows also how some results can be framed within
the theory of continued fractions. The formulas that allow the
explicit determination of the period are given in Section 5, both
for the rational and irrational cases. Finally, some conclusions and
further work are outlined in Section 6.

\section{The model}\label{Section:Model}

The model we analyze in this paper is given by the following set of
three coupled first order ODEs:
\begin{equation}
\dot{z}_{n}+i\,\omega \,z_{n}=\frac{g_{n+2}}{z_{n}-z_{n+1}}+\frac{g_{n+1}}{%
z_{n}-z_{n+2}}~.  \label{EqMot}
\end{equation}%

\smallskip

\noindent \textit{Notation}: Here and hereafter indices such as $n$,
$m$ range from $1$ to $3$ and are defined $\tmodulus{(3)}$, where
$\tmodulus$ is defined for all positive integers $a$ and $b$ as
follows
\begin{equation}\label{tmodulus}
   a \tmodulus{(b)} = \left \{
                         \begin{array}{cc}
                            a \modulus{(b)} & \mbox{ if }\,\, a \modulus{(b)} \neq 0 \\
                            b & \mbox{ if }\,\, a \modulus{(b)} = 0 \\
                         \end{array}
                \right. \mbox{ . }
\end{equation}
\noindent (hence, for instance, $3\tmodulus{(3)}=3$,
$4\tmodulus{(3)}=1$, the usefulness of this
notation will become apparent below). The dependent variables
$z_n=z_n(t)$ are complex functions and indicate the
positions of three interacting bodies in the
plane; the independent variable $t$ (``physical time'') is
real and the superimposed dots denote differentiation with
respect to $t$. The parameter $\omega$ is strictly positive and is
associated to the period
\begin{equation}\label{T}
T=\frac{\pi}{\omega}\,.
\end{equation}

\noindent We fix hereafter $\omega=\pi$ without loss of generality so that the fundamental period $T=1$.

\noindent The three quantities $g_{n}$ are arbitrary ``coupling constants'' (possibly also complex; but in this
paper we restrict consideration only to the case with
real coupling constants).

\smallskip

In the following we will focus only on the ``semisymmetrical case''
characterized by the equality of two of the three
coupling constants, say
\begin{equation}
g_{1}=g_{2}=g\,,\qquad g_{3}=f\,,\label{Symm}
\end{equation}%
\noindent since in this case the treatment is simpler yet still
adequate to exhibit most aspects of the phenomenology we are
interested in (see \cite{CGSS2008-I}). In this semisymmetrical case
it is convenient to introduce the constant $\mu$,
\begin{equation}
\mu =\frac{f+2\,g}{f+8\,g} \label{mu}
\end{equation}%
\noindent whose value, as we shall see, plays an important role in
determining the dynamical evolution of our model.

In \cite{CGSS2008-I} the general solution of system (\ref{EqMot})
was given. In the semisymmetrical case (\ref{Symm}), if $\mu\neq 0$
and $\mu\neq 1$, the solution of (\ref{EqMot}) reads as follows.

\begin{subequations}
\begin{flalign}\quad
z_{s}(t)=&Z\,{\rm e}^{i\pi t} -\frac{%
2\,z_{3}(0)-z_{1}(0)-z_{2}(0)}{6\,\sqrt{\mu }}\,\left[ \eta \,\exp
\left(
-2\,i\,\pi \,t\right) +1\right] ^{\,1/2}\cdot  \nonumber \\
\qquad\quad &\cdot \left( -\left[ \check{w}\left( t\right)
\right] ^{\,1/2}+\left( -1\right) ^{s}\,\left[ 12\,\mu
-3\,\check{w}\left( t\right) \right] ^{\,1/2}\right) ,\qquad s=1,2,
\label{SolSim-a}
\end{flalign}%
\begin{flalign}
z_{3}\left( t\right) =Z\,{\rm e}^{i\pi t} -\frac{%
2\,z_{3}(0)-z_{1}(0)-z_{2}(0)}{3\,\sqrt{\mu }}\,\left[ \eta \,\exp
\left(
-2\,i\,\pi \,t\right) +1\right] ^{\,1/2}\,\left[ \check{w}\left( t\right) %
\right] ^{\,1/2},  \label{SolSim-b}
\end{flalign}%
\end{subequations}
\noindent where
\begin{equation}
Z=\frac{z_{1}+z_{2}+z_{3}}{3}%
\label{centerofmass}
\end{equation}

\noindent is the center of mass, $\check{w}\left( t\right)$ is the
solution of the nondifferential equation
\begin{equation}
\left[ \check{w}\left( t\right) -1\right] ^{\,\mu -1}\,\left[ \check{w}%
\left( t\right) \right] ^{\,-\mu }=R\,\exp \left( 2\,i\,\pi \,t\right) +%
\bar{\xi}=R\,\left[ \exp \left( 2\,i\,\pi \,t\right) +\eta
\right] , \label{Eqwtilde}
\end{equation}%

\noindent and the constants $R$, $\bar{\xi}$ and $\eta$ are defined
in terms of the initial data:
\begin{subequations}
\begin{flalign}
\eta &=\frac{i\,\pi \,\left\{ \left[ z_{1}(0)-z_{2}(0)\right]
^{\,2}+\left[ z_{2}(0)-z_{3}(0)\right] ^{\,2}+\left[
z_{3}(0)-z_{1}(0)\right] ^{\,2}\right\} }{3\,\left( f+2\,g\right)}-1\,, \label{Constants-eta}
\\
R&=\frac{3\,\left( f+8\,g\right) }{2\,i\,\pi \,\left[ 2%
\,z_{3}(0)-z_{1}(0)-z_{2}(0)\right] ^{\,2}}\,\left[ 1-\kappa \right]
^{\,\mu -1}\,,  \label{Constants-R}
\\
\bar{\xi}&=R\,\eta\,,  \label{Constants-xibar}
\end{flalign}%

\noindent where $\kappa$ in (\ref{Constants-R}) is given by
\begin{equation}
\kappa =\frac{2\,\mu \,\left[ 2\,z_{3}(0)-z_{1}(0)-z_{2}(0)%
\right]^{\,2}}{\left[z_{1}(0)-z_{2}(0)\right]^{\,2}+\left[
z_{2}(0)-z_{3}(0)\right]
^{\,2}+\left[z_{3}(0)-z_{1}(0)\right]^{\,2}}\,. \label{Constants-k}
\end{equation}%
\end{subequations}

\noindent At this stage, it is mandatory to note that
(\ref{Constants-R}) contains a degeneracy of order $\mu-1$. In the
following sections we will explain the role of such a degeneracy in
the construction of the solution and how to remove it.

If we now set
\begin{subequations}
\begin{align}
&\check{w}(t)\equiv w\left[ \xi \left( t\right) \right]\,,
\label{wtilde}\\
&\xi =R\,\exp \left( 2\,i\,\pi \,t\right) +\bar{\xi}=R\,\left[
\exp \left( 2\,i\,\pi \,t\right) +\eta \right] ,  \label{ksi}
\end{align}%
\end{subequations}
\noindent we can rephrase (\ref{Eqwtilde}) as follows
\begin{equation}
\left[ w\left( \xi \right) -1\right] ^{\,\mu -1}\,\left[ w\left( \xi \right) %
\right] ^{\,-\mu }=\xi \,.  \label{Eqw}
\end{equation}%
Note that this equation is independent of the initial data;
it only features the constant $\mu$, which only depends on the
coupling constants, see (\ref{mu}). Moreover, it defines the Riemann
surface $\Gamma$ consisting of points $(\xi,w)\in\Gamma$ such that
(\ref{Eqw}) is satisfied.

As $\xi$ travels in the complex $\xi$-plane on the circle $\Xi$ defined by
(\ref{ksi}), the dependent variable $w\left( \xi \right)$ travels on
the Riemann surface determined by its dependence on the
complex variable $\xi$, as entailed by the equation
(\ref{Eqw}) that relates $ w\left( \xi \right)$ to its argument $\xi
$ -- starting at $t=0$ from $\xi =\xi_{0}$,
\begin{subequations}
\begin{equation}
\xi _{0}=\bar{\xi}+R=\left( \eta +1\right)\,R\,,
\label{InitData-ksi}
\end{equation}%
(see (\ref{Constants-eta})-(\ref{Constants-k})) and correspondingly
from $w(\xi_{0})=w_{0}$,
\begin{equation}
w_{0}=\frac{1}{\kappa}=\frac{\left[z_{1}(0)-z_{2}(0)\right]^{\,2}+
\left[z_{2}(0)-z_{3}(0)\right]^{\,2}+
\left[z_{3}(0)-z_{1}(0)\right]^{\,2}}{2\,\mu\,\left[2\,z_{3}(0)-z_{1}(0)-z_{2}(0)\right]^{\,2}}\,.
\label{InitData-w}
\end{equation}%
\end{subequations}

\noindent We observe that $\xi$ and $\xi_{0}$ feature the same
degeneracy as $R$.

We still must fix the degeneracy of the square roots appearing in
(\ref{SolSim-a})-(\ref{SolSim-b}). Since there is no degeneracy in
the determination of $w_{0}$ and $\eta$, the determination of the
signs of the square roots in (\ref{SolSim-a})-(\ref{SolSim-b}) is
fixed by demanding that these formulae are consistent with the initial data at $t=0$.

Despite the periodicity in time of $\xi$, the corresponding time
evolution of $w$ via \eqref{Eqw} could be much more complicated
(possibly aperiodic) due to the multivaluedness of $w$ as a function
of $\xi$. This evolution must be studied by lifting a circular path
to the Riemann surface defined by \eqref{Eqw}. This motivates the
necessity to study the geometric structure of the Riemann surface,
which we address in the following section.

\section{The Riemann surface}\label{Sec:RiemannSurface}

In this section we discuss the structure of the Riemann surface
$\Gamma$ consisting of points $(\xi,w)\in\Gamma$ such that
(\ref{Eqw}) is satisfied.

For rational values of $\mu=p/q$, which is the case if both
the coupling constants $f$ and $g$ are rational numbers,
see (\ref{mu}), equation (\ref{Eqw}) describes an algebraic
curve. We treat the case of irrational $\mu$ via an
appropriate limit of the case with rational $\mu$.
Moreover, as already pointed out in \cite{CGSS2008-I} and as we
recall in the next section, one needs to distinguish among the three
cases with $\mu<0$, $0<\mu<1$ and $\mu>1$.

We start the analysis of the structure of $\Gamma$ by ignoring the
dependence of $\xi$ on the ``physical'' time $t$ (see (\ref{ksi})),
 and by considering instead $\xi$ as an
independent complex variable, evolving on a
generic (possibly closed) path on the complex
$\xi$-plane; only after having thereby obtained an appropriate
understanding of the topological properties of $\Gamma$, we proceed
and analyze the consequences of the ``physical movement'' of $\xi$
along the circle $\Xi$, as described by (\ref{ksi}).

\subsection{Movable and fixed singularities}\label{Sec:Singularities}

The function $w(\xi)$ defined by (\ref{Eqw}) features two types of
singularities: the ``fixed'' ones occurring at values of the
independent variable $\xi$ -- and correspondingly of the dependent
variable $w$ -- that can be read directly from the structure of
(\ref{Eqw}); and the ``movable'' ones occurring at values of the
independent and dependent variables, $\xi $ and $w$, that cannot be
directly read from the structure of (\ref{Eqw}) (they ``move'' as
the initial data are modified).

In order to investigate the nature of the movable singularities, it
is convenient to differentiate (\ref{Eqw}), thereby obtaining (by
repeated use of (\ref{Eqw}))
\begin{equation}
\xi \,w^{\prime }=-\frac{w\,\left(w-1\right)}{w-\mu},
\label{EqDiffw}
\end{equation}%
\noindent where the prime indicates differentiation with respect to
$\xi$. The position of the movable singularities,
$\xi_{\mathrm{b}}$, and the corresponding values of the dependent
variable, $w_{b}\equiv w(\xi_{\mathrm{b}})$, are then characterized
by the vanishing of the denominator in the right-hand side of this
formula, yielding the relation
\begin{subequations}\label{bpscircle}
\begin{equation}
w_{b}=\mu,  \label{LocMovableSing-w}
\end{equation}%

\noindent which, combined with (\ref{Eqw}) (at $\xi =\xi
_{\mathrm{b}})$ is easily seen to yield
\begin{equation}
\xi _{\mathrm{b}}=\xi _{\mathrm{b}}^{(k)}=r\,\exp \left( 2\,\pi
\,i\,\mu \,k\right)\, , \qquad k=1,2,3,\dots  \label{LocMovableSing-ksi}
\end{equation}%
\begin{equation}
r=\left( \mu -1\right) ^{\,-1}\,\left( \frac{\mu -1}{\mu }\right)
^{\,\mu }\,. \label{LocMovableSing-r}
\end{equation}%
\end{subequations}

\noindent In (\ref{LocMovableSing-r}) it is understood that the
principal determination is to be taken of the $\mu$-th power
appearing in the right-hand side. Formula (\ref{LocMovableSing-ksi})
shows clearly that the number of these branch points is infinite if
the parameter $\mu$ is irrational ($\mu \notin \mathbb{Q}$), and
that they then sit densely on the circle $B$ in the complex
$\xi$-plane centered at the origin and having radius $r$, see
(\ref{LocMovableSing-r}). On the contrary, if $\mu$ is rational
($\mu \in \mathbb{Q}$) the branch points sit again on the circle $B$
in the complex $\xi$-plane, but there are only a finite number of
them. As proved in \cite{CGSS2008-I}, these movable singularities
are all square-root branch points.

Then, let us consider the ``fixed'' singularities, which clearly can
only occur at $\xi =\infty $ and at $\xi =0$ with corresponding
values for $w$.

Two behaviors of $w(\xi)$ are possible for $\xi \approx
\infty$, depending on the value of (the real part of) $\mu$. The
first is characterized by the \textit{ansatz}
\begin{subequations}
\begin{equation}
w(\xi )=a\!\,\xi ^{\!\beta }+o\left( \left\vert \xi \right\vert ^{\beta}\right) \,,\qquad \beta<0,  \label{Ansatz1a}
\end{equation}%
\noindent and its insertion in (\ref{Eqw}) yields%
\begin{equation}
\beta =-\frac{1}{\mu }\,,\qquad a^{\mu }=-\exp (i\,\pi \,\mu ),
\label{Ansatz1b}
\end{equation}%
\noindent which is consistent with (\ref{Ansatz1a}) iff%
\begin{equation}
\mu>0\,.  \label{Ansatz1c}
\end{equation}%
\end{subequations}
\noindent The second is characterized by the \textit{ansatz}
\begin{subequations}
\begin{equation}
w(\xi )=1+a\!\,\xi ^{\!\beta }+o\left( \left\vert \xi \right\vert
^{\beta}\right) \,,\qquad \beta<0\,, \label{Ansatz2a}
\end{equation}%
\noindent and its insertion in (\ref{Eqw}) yields%
\begin{equation}
\beta =\frac{1}{\mu -1}\,,\qquad a^{\!\mu -1}=1,  \label{Ansatz2b}
\end{equation}%
which is consistent with (\ref{Ansatz2a}) iff%
\begin{equation}
\mu<1~.  \label{Ansatz2c}
\end{equation}%
\end{subequations}

\noindent We therefore conclude that there are only three
possibilities:

\begin{itemize}
\item if $\mu>1$, only the first \textit{ansatz}, (\ref{Ansatz1a})-(\ref{Ansatz1c}),
is applicable, and it characterizes the nature of the branch point
of $w(\xi )$ at $\xi
=\infty$;%
\item if $\mu<0$, only the second \textit{ansatz}, (\ref{Ansatz2a})-(\ref{Ansatz2c}),
is applicable, and it characterizes the nature of the branch point
of $w(\xi)$ at $\xi =\infty$;%
\item if $0<\mu<1$, both \textit{ans\"{a}tze},
(\ref{Ansatz1a})-(\ref{Ansatz1c}) and
(\ref{Ansatz2a})-(\ref{Ansatz2c}), are applicable, so both types of
branch points occur at $\xi =\infty$.
\end{itemize}
The special cases $\mu=0$ and $\mu=1$ require a separate treatment \cite{sommacal}.
Next, let us investigate the nature of the singularity at $\xi =0$.
Two behaviors are possible: either
\begin{subequations}
\begin{equation}
w(\xi )=a\!\,\xi ^{\,\beta }+o\left( \left\vert \xi \right\vert ^{\beta}\right) \,,\qquad \beta>0,  \label{Ansatz3a}
\end{equation}%
\begin{equation}
\beta =-\frac{1}{\mu }\,,\qquad a^{\,\mu }=-\exp \left( i\,\pi \,\mu
\right) , \label{Ansatz3b}
\end{equation}%
\noindent which is applicable if and only if%
\begin{equation}
\mu<0~;  \label{Ansatz3c}
\end{equation}%
\end{subequations}
\noindent or
\begin{subequations}
\begin{equation}
w(\xi )=1+a\!\,\xi ^{\,\beta }+o\left( \left\vert \xi \right\vert ^{\beta}\right) \,,\qquad \beta>0,  \label{Ansatz4a}
\end{equation}%
\begin{equation}
\beta =\frac{1}{\mu -1}\,,\qquad a^{\,\mu -1}=1,  \label{Ansatz4b}
\end{equation}%
\noindent which is applicable if and only if
\begin{equation}
\mu>1~.  \label{Ansatz4c}
\end{equation}%
\end{subequations}

\noindent This analysis shows that the function $w(\xi)$ features a
branch point at $\xi =0$ the nature of which is
characterized by the relevant exponent $\beta$, see (\ref{Ansatz3b})
or (\ref{Ansatz4b}), whichever is applicable (see (\ref{Ansatz3c})
and (\ref{Ansatz4c})). There is no branch point at
all at $\xi =0$ if neither one of the two inequalities
(\ref{Ansatz3c}) and (\ref{Ansatz4c}) holds, namely if
$0<\mu<1$.

Moreover, we observe that the case $\mu<0$ can be immediately worked
out from the case $\mu>1$ via the following replacement
\begin{equation}\label{replacement}
w\mapsto w-1\,,\qquad \xi\mapsto -\xi\,,\qquad \mu\mapsto 1-\mu\,.
\end{equation}

\noindent For this reason, only the two cases $0<\mu<1$ and $\mu>1$
lead to different Riemann surfaces. Although the geometrical
properties of the Riemann surface are quite different in the two
cases, the techniques employed in their study are essentially the
same. For this reason, to avoid unnecessary repetitions we
concentrate in this paper only on the first case  $0<\mu<1$, which,
as we shall see, leads to a rich and complex behavior. The second
case $\mu>1$ which also has interesting consequences for the
dynamics, specially in the case $\mu\not\in\mathbb Q$, shall be
discussed in a subsequent paper.

\subsection{The case $0<\mu<1$ and $\mu\in\mathbb{Q}$}\label{Sec:mu-0<mu<1-rational}%

If $\mu$ is a rational number in the open interval $(0,1)$,
namely if
\begin{equation}\label{mu-0<mu<1-rational}
\mu\in\mathbb{Q} \qquad \mbox{and} \qquad 0<\mu=\frac{p}{q}<1 \,\,,
\end{equation}

\noindent where $p\in\mathbb{N}$ and $q\in\mathbb{N}^{+}$ are
coprime natural numbers, then (\ref{Eqw}) implies that $\Gamma$ is
an algebraic Riemann surface characterized by the
polynomial equation
\begin{equation}
\label{algebraic curve 0<p<q} (w-1)^{q-p}\,w^{p}\,\xi^q=1 \qquad
\mbox{with} \qquad 0 < p < q \,\,,
\end{equation}

\noindent which defines the $q$-valued function $w=w(\xi)$. Since,
$\forall\xi\in\mathbb{C}$, the polynomial (\ref{algebraic curve
0<p<q}) admits $q$ complex roots and each root corresponds to a
sheet of the Riemann surface $\Gamma$, it follows that $\Gamma$ is a
$q$-sheeted covering of the complex $\xi$-plane.

Condition (\ref{mu-0<mu<1-rational}), via formulae
(\ref{Ansatz1a})-(\ref{Ansatz1c}) and
(\ref{Ansatz2a})-(\ref{Ansatz2c}), entails that, for $\xi \approx
\infty$, the $\infty$-\textit{configuration} of the $q$ roots of
(\ref{algebraic curve 0<p<q}) consists of $p$ roots lying on a small
circle of radius $O(|\xi |^{-\frac{q}{p}})$ around the origin, and
of  $(q-p)$ roots lying on a small circle of radius $O(|\xi
|^{-\frac{q}{q-p}})$ around $1$ (see Figure \ref{Fig:Basic
configuration p5 q12}). From the point of view of the Riemann
surface, we see that, at $\xi=\infty$, the branch point
$(\infty,0)$, of order $(p-1)$, connects $p$ sheets and the branch
point $(\infty,1)$, of order $(q-p-1)$, connects $(q-p)$ sheets.

\bigskip
\begin{figure}[h]
\centering
\fbox{\includegraphics[width=8cm]{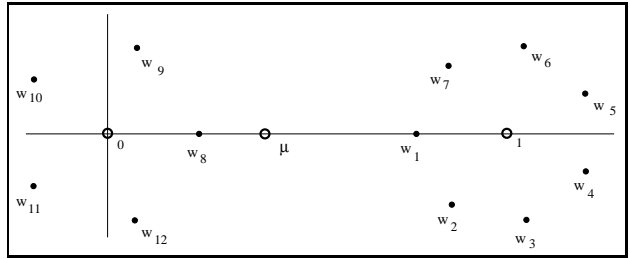}}
\caption{The $\infty$-\textit{configuration} of the $q$ roots of
(\ref{algebraic curve 0<p<q}), for $q=12$ and $p=5$. The labeling of
the roots is explained in the text.}\label{Fig:Basic configuration
p5 q12}
\end{figure}
\bigskip

In the finite part of $\Gamma$, condition (\ref{mu-0<mu<1-rational})
and formulae (\ref{LocMovableSing-w})-(\ref{LocMovableSing-r}) imply
that there are only $q$ square-root branch points:
\begin{subequations}
\begin{equation} \label{SRBP1 0<mu<1}
(\xi^{(j)}_\mathrm{b},\mu)\in \Gamma\,,\qquad j=1,...,q\,\,,
\end{equation}
\noindent defined by the equation:
\begin{equation} \label{SRBP2 0<mu<1}
\xi^q=\frac{(-)^{q-p}q^q}{p^p(q-p)^{q-p}}\,\,.
\end{equation}

\noindent These $q$ square-root branch points correspond to the
collision of a pair of roots of (\ref{algebraic curve 0<p<q}); they
are clearly located on the circle $B$ in the complex $\xi$-plane
centered at the origin and having radius $r_b$
\begin{equation}\label{B circle radius 0<mu<1 mu rational}
r_b=\frac{q}{q-p}\left(\frac{q-p}{p}\right)^\frac{p}{q}>0.
\end{equation}
\end{subequations}

\begin{figure}[h]
\centering
\fbox{\includegraphics[width=8cm]{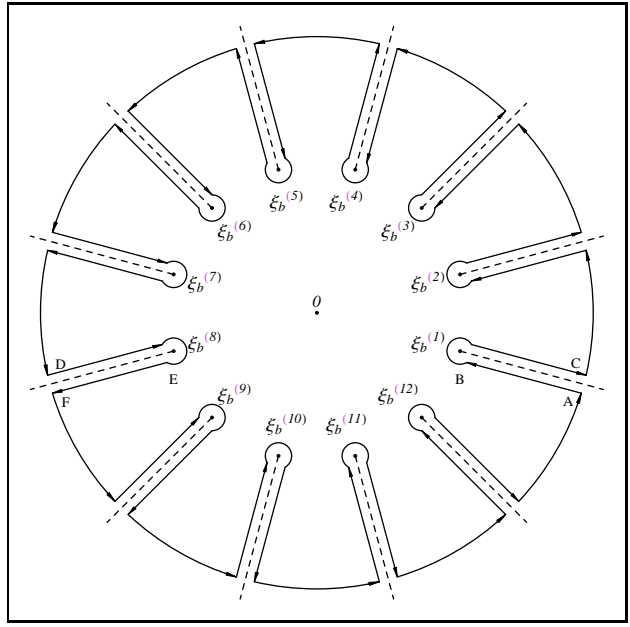}}%
\caption{The cut $\xi$-plane of the Riemann surface $\Gamma$ with
$q=12$ and $p=5$ is the interior of the oriented contour
$\gamma$.}\label{Fig:Cut plane p5 q12}
\end{figure}

\noindent It is convenient to sort these $q$ square-root branch
points sequentially in counterclockwise order (see Figure
\ref{Fig:Cut plane p5 q12}); an opportune choice of the
first square-root branch point $\xi^{(1)}_\mathrm{b}$,
clearly arbitrary at this stage, is suggested by the direct problem
and will be discussed later.

\bigskip

The genus of $\Gamma$ is $0$; this is an immediate consequence of
the Hurwitz formula, $V=2(J+G-1)$, where $V$ is the ramification
index of the surface, $J$ is the number of sheets and $G$ is its
genus. In our case: $J=q$ and $V=q+(q-p-1)+(p-1)=2(q-1)$, entailing
$G=0$.

Moreover, equation (\ref{algebraic curve 0<p<q}) exhibits several
symmetries. The ones used below are:

\begin{itemize}
    \item {\it Symmetry under a $2\pi/q$ rotation}. The set of roots is
          left invariant by a rotation around the origin of the $\xi$-plane by an
          angle $2\pi/q$ (and of course any integer multiple of it).
    \item {\it Symmetry along the cuts}. If $\xi$ belongs to the rays passing through the square-root branch points, (\ref{SRBP2 0<mu<1})
          implies that (\ref{algebraic curve 0<p<q}) is a polynomial
          equation featuring real coefficients. Therefore, its roots
          are real or in complex conjugate pairs.
          Consider the semiline cuts
          $\gamma_j$, $j=1,...,q$, defined by
          \begin{equation}\label{Cuts 0<mu<1}
          \gamma_{j}=\{\xi,\,\arg{\xi}=\arg{\xi^{(j)}_\mathrm{b}},\,|\xi |\ge
          |\xi^{(j)}_\mathrm{b} |\}\,,\qquad j=1,...,q\,.
          \end{equation}
          Then, if $\xi\in\gamma_{j}$, $j=1,...,q$, two of the $q$ roots lie on
          the segment $(0,1)$.  If, in addition, $\xi\approx\infty$, then one of
          these two roots belongs to the small circle around the origin and
          the other one belongs to the small circle around $1$ (see Figure \ref{Fig:Basic configuration p5 q12},
          where the $\infty$-\textit{configuration} of the $q$ roots is shown for $\xi\approx\infty$ and $\xi\in\gamma_1$).%
\end{itemize}

\subsection{Roots dynamics and topological properties}\label{Sec:mu-0<mu<1-rational:Topological}%

In order to understand the topological properties of the Riemann
surface $\Gamma$, we first cut the $\xi$-complex plane along the
rays $\gamma_{j}$, $j=1,...,q$, see (\ref{Cuts 0<mu<1}), from the
square root branch points $\xi^{(j)}_\mathrm{b}$ to the branch point
at $\infty$, and we introduce the closed contour $\gamma$ whose
interior is the cut $\xi$-plane thereby obtained (see Figure
\ref{Fig:Cut plane p5 q12}). Let $\mathcal{F}_j$ be the Riemann
sheet of $\Gamma$ associated to the root $w_j$. Our goal is to
construct the $q$ images $\mathcal{I}_{j}$, $j=1,...,q$, of the cut
$\xi$-plane, corresponding to the $q$ roots $w_{j}(\xi)$ of
(\ref{algebraic curve 0<p<q}), $j=1,...,q$, and to study their
connections.

\smallskip

\noindent \textit{Notation}: Here and hereafter, through Sections
\ref{Sec:mu-0<mu<1-rational} and \ref{Sec:mu-0<mu<1-irrational}, all
the indices in the designation of the square-root branch points
$\xi^{(j)}_\mathrm{b}$, the roots $w_j$, the images
$\mathcal{I}_{j}$ and the sheets $\mathcal{F}_j$ are defined
$\tmodulus{(q)}$ via the congruence defined in (\ref{tmodulus}),
namely
$\xi^{(j)}_\mathrm{b}\equiv\xi^{(j\footnotesize{\tmodulus}{(q)})}_\mathrm{b}$
or
$\mathcal{F}_j\equiv\mathcal{F}_{j\footnotesize{\tmodulus}{(q)}}$.

\smallskip

Our strategy will be to start from the
$\infty$-\textit{configuration}, namely the root configuration
entailed by setting $\xi\approx\infty$, and to map out the structure
of the Riemann surface as $\xi$ is moved in along rays and it is
moved around along circles.

Starting with $\xi\in\gamma_{j}$, it is natural to call $w_{j}$ the
root lying on the segment $(0,1)$ and  belonging to the small circle
around $1$. Then $w_{j+1},...,w_{j+q-p-1}$ are the other roots of
this circle, enumerated sequentially in counterclockwise order.
Analogously, we denote by $w_{j+q-p}$ the root lying on the segment
$(0,1)$ and belonging to the small circle around $0$; and by
$w_{j+q-p+1},...,w_{j+q-1}$ the other roots of this circle,
enumerated sequentially in counterclockwise order (see Figure
\ref{Fig:Basic configuration p5 q12}).

Using the large $\xi$ asymptotics, (\ref{Ansatz1a})-(\ref{Ansatz1c})
and (\ref{Ansatz2a})-(\ref{Ansatz2c}), and the symmetries of
$\Gamma$ described above, we infer the following basic motions.

\begin{figure}[h]
\centering
\fbox{\includegraphics[width=8cm]{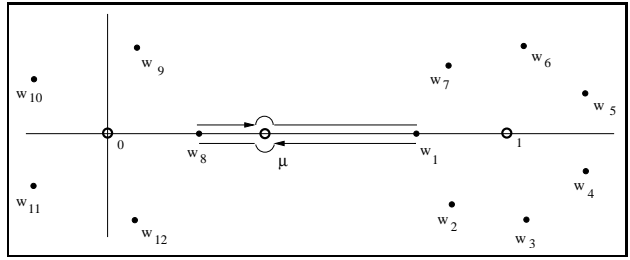}}
\caption{Exchange of a pair of roots. As $\xi$ travels on the
contour surrounding the cut $\gamma_{j}$ (for example moving from
the point $A$ to the point $C$ along the path shown in Figure
\ref{Fig:Cut plane p5 q12}), $w_{j}$ and $w_{j+q-p}$ interchange
their positions, while the other $(q-p)$ roots are essentially
unaffected by this motion.}\label{Fig:Roots exchange p5 q12}
\end{figure}

\begin{figure}[h]
\centering \fbox{\includegraphics[width=8cm]{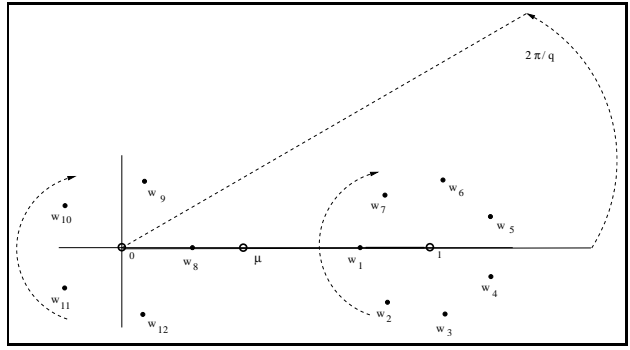}}
\caption{Cyclic permutation of the two groups of roots. If $\xi\in
\gamma_j$, the roots $w_{j}$ and $w_{j+q-p}$ are aligned on the
segment $(0,1)$; after a $2\pi/q$ counterclockwise rotation of
$\xi$, from the cut $\gamma_j$ to the cut $\gamma_{j+1}$, the two
groups of roots $\{w_j,..,w_{j+q-p-1}\}$ and
$\{w_{j+q-p-1},..,w_{j+q-1}\}$ undergo a clockwise rotation. When
$\xi$ reaches $\gamma_{j+1}$, then the roots $w_{j+1}$ and
$w_{j+1+q-p}$ get aligned on the segment
$(0,1)$.}\label{Fig:Rotation p5 q12}
\end{figure}

\begin{itemize}
\item As $\xi$ moves along the cut $\gamma_{j}$,
    from $\infty$ to $\xi^{(j)}_{\mathrm{b}}$, the two roots $w_{j}$ and
    $w_{j+q-p}$, lying on the segment $(0,1)$, move along it, from the
    small circles around $0$ and $1$ to the collision point $\mu$. In
    addition, being $\xi^{(j)}_{\mathrm{b}}$ a branch point of square-root type,
    a $2\pi$ rotation of $\xi$ around it corresponds to a $\pi$ rotation
    of these two roots around $\mu$. All this implies that, if $\xi$
    travels along the contour surrounding the cut $\gamma_{j}$ (for
    instance, moving from the point $A$ to the point $C$ along the path
    shown in Figure \ref{Fig:Cut plane p5 q12}), then the two roots
    $w_{j}$ and $w_{j+q-p}$ involved in the collision {\it exchange
    their position} (see Figure \ref{Fig:Roots exchange p5 q12}). The
    remaining roots are essentially unaffected by this motion, moving
    back and forth on lines and going back to their starting positions.
    We have thereby established the first basic motion: %
    \smallskip%
    \textit{a motion of $\xi$ around the branch cut $\gamma_j$ yields an exchange of the two
    roots $w_j$ and  $w_{j+q-p}$}.
    \smallskip%
\item If, starting from the cut $\gamma_{j}$, $\xi$ performs a $2\pi/q$ counterclockwise rotation,
    moving from the cut $\gamma_{j}$ to the cut $\gamma_{j+1}$, then the $(q-p)$ roots
    surrounding $1$ undergo a clockwise rotation around $1$,
    while the $p$ roots surrounding $0$ undergo a clockwise rotation
    around $0$. When $\xi$, starting from $\gamma_j$, reaches $\gamma_{j+1}$, two new roots
    belonging to the two small circles get aligned on the segment
    $(0,1)$; they are just $w_{j+1}$ and $w_{j+1+q-p}$ (see
    Figure \ref{Fig:Rotation p5 q12}). Therefore the two sets of roots undergo {\it cyclic
    permutations}, which is the second basic motion: %
    \smallskip%
    \textit{a rotation of $\xi$ from $\gamma_j$ to $\gamma_{j+1}$ produces a
    cyclic permutation of the two sets of roots in the $\infty$-\textrm{configuration}
    $\left\{w_1,...,w_{q-p}\right\}$ and $\left\{w_{q-p+1},...,w_{q}\right\}$}.
    \smallskip%
\end{itemize}

\begin{figure}[h]
\centering \fbox{\includegraphics[width=8cm]{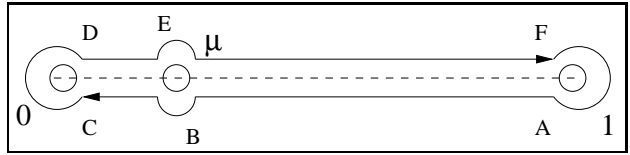}}
\caption{As $\xi$ travels on the closed contour $\gamma$ in Figure
\ref{Fig:Cut plane p5 q12}, each root travels on a closed contour
around the cut $[0,1]$ in the $w$-plane. Therefore the image of the
cut $\xi$-plane is the $w$-plane cut along the segment
$[0,1])$.}\label{Fig:01Cut p5 q12}
\end{figure}

\noindent Repeating $q$ times the above two motions with respect to
the other cuts, in sequential order, the point $\xi$ draws the whole
closed contour $\gamma$ and correspondingly, due to the
above-mentioned basic motions and symmetries, each root $w_{j}$
draws the closed contour in Figure \ref{Fig:01Cut p5 q12} around the
cut $[0,1]$ of the $w$-plane. The following details can be given
(compare Figure \ref{Fig:Cut plane p5 q12} and Figure \ref{Fig:01Cut
p5 q12}).

\begin{enumerate}
\item As $\xi$ moves on $\gamma$ through the points $A$, $B$
      and $C$, around the cut $\gamma_{j}$, $w_{j}$ moves along the cut
      $[0,1]$ through the homologous points, exchanging its position
      with the root $w_{j+q-p}$.
\item As $\xi$ moves on $\gamma$ from $C$ to $D$, the
      relevant part of the motion of $w_{j}$ consists in a clockwise
      rotation around $0$, from $C$ to $D$.
\item As $\xi$ moves on $\gamma$ through the points
      $D$, $E$ and $F$, around the cut $\gamma_{j-q+p}$, $w_{j}$
      moves along the cut $[0,1]$ through the homologous points,
      exchanging its position with the root $w_{j-q+p}$.
\item As $\xi$ completes the contour $\gamma$, moving from
      $F$ to the starting point $A$, also $w_{j}$ completes its closed
      contour around $[0,1]$, and the relevant part of this motion
      consists in a clockwise rotation around $0$, from $F$ to $A$.
\end{enumerate}

From the above considerations we finally infer the following

\smallskip

\noindent \textbf{Topological properties of $\Gamma$ for
$\mathbf{\mu=p/q}$ and $\mathbf{0<p<q}$.} {\it The Riemann surface
$\Gamma$ defined in (\ref{algebraic curve 0<p<q}) is a $q$-sheeted
covering of the $\xi$-plane of genus $0$. In the finite part of
$\Gamma$ there are $q$ square-root branch points:
$(\xi^{(j)}_\mathrm{b},\mu)\in\Gamma$, $j=1,...,q$ (see (\ref{Cuts
0<mu<1})). If $\mathcal{F}_j$ is the sheet associated to the root
$w_j$, then the $j$-th branch point connects the sheets
$\mathcal{F}_{j}$ and $\mathcal{F}_{j+q-p}$. Each sheet
$\mathcal{F}_j$ contains just the two square-root branch points
$(\xi^{(j)}_\mathrm{b},\mu)$ and $(\xi^{(j-q+p)}_\mathrm{b},\mu)$.
The compactification of $\Gamma$ is achieved at $\xi=\infty$, where
the branch point $(\infty,1)$, of order $(q-p-1)$, connects the
first $(q-p)$ sheets, and where the branch point $(\infty,0)$, of
order $p-1$, connects the remaining $p$ sheets. As $\xi$ turns
counterclockwise around $(\infty,1)$, the connected sheets are
visited in the order: $\mathcal{F}_{j}$, $\mathcal{F}_{j+p}$,
$\mathcal{F}_{j+2p}$,...; as $\xi$ turns counterclockwise around
$(\infty,0)$ instead, the connected sheets are visited in the order:
$\mathcal{F}_{j}$, $\mathcal{F}_{j+q-p}$,
$\mathcal{F}_{j+2(q-p)}$,...}
\smallskip

\subsection{The physical root}\label{Sec:mu-0<mu<1-rational:Physical Root}%

Once a labeling system has been introduced for the square-root
branch points $\xi^{(j)}_\mathrm{b}$'s and for the roots $w_j$'s, in
order to solve our original problem (\ref{EqMot}) we need to
indicate which of the $q$ roots of (\ref{algebraic curve 0<p<q})
corresponds to the physical root $\check{w}(t)\equiv w(\xi)$
appearing in (\ref{SolSim-a})-(\ref{SolSim-b}) and satisfying
(\ref{Eqwtilde}).

As implied by (\ref{InitData-ksi}) and (\ref{InitData-w}), at $t=0$
we have that $\xi=\xi_0$ and one of the $q$ roots of (\ref{algebraic
curve 0<p<q}) is $w_0$. Therefore, it is sufficient to label $w_0$
in order to label the physical root. Observing (\ref{InitData-ksi})
and (\ref{Constants-R}), we immediately note that $\xi_0$ is a
$q$-valued function of the initial conditions $z_{1}(0)$, $z_{2}(0)$
and $z_{3}(0)$; on the contrary, from (\ref{InitData-w}) we infer
that $w_0$ is a single-valued function of the initial conditions
$z_{1}(0)$, $z_{2}(0)$ and $z_{3}(0)$. Via (\ref{Constants-xibar})
and (\ref{ksi}), we get that $\xi_0$ and $\bar{\xi}$, the center of
the circle $\Xi$ on which the $\xi$ variable moves, have the same
degeneracy. This degeneracy can be removed by arbitrarily fixing one
of the $q$ determinations of $R$ in (\ref{Constants-R}), placing the
circle $\Xi$ on the $\xi$-plane and then labeling counterclockwise
the square-root branch points lying on the circle $B$ defined by
(\ref{SRBP2 0<mu<1}) so that $\xi^{(1)}_\mathrm{b}$ is the first
branch point encountered by moving counterclockwise on the arc of
the circle $B$ obtained by intersecting the two circles and
contained inside the circle $\Xi$ (see Figure \ref{Fig:circles -
0<mu<1}). When the two circles $\Xi$ and $B$ do not intersect, then
the first branch point $\xi^{(1)}_\mathrm{b}$ on $B$ can be chosen
arbitrarily.

\begin{figure}[h]
\centering \fbox{\includegraphics[width=8cm]{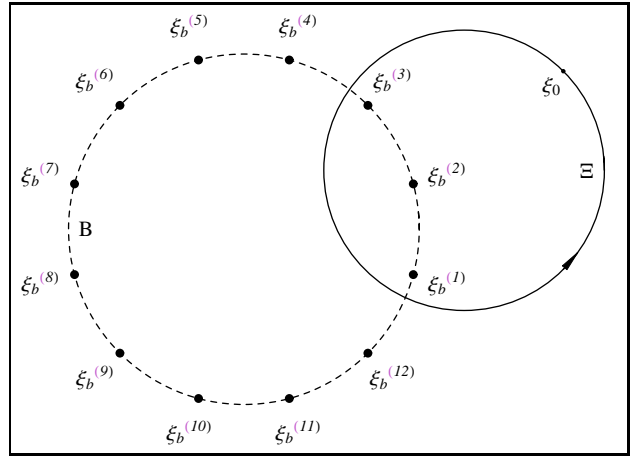}}
\caption{The variable $\xi$ moves counterclockwise on the circle
$\Xi$ starting from $\xi_0$. The first square-root branch point on
the circle $B$ is the first one encountered by moving
counterclockwise on the arc of $B$, obtained by intersecting the two
circles and contained inside the circle $\Xi$. In this example $p=5$ and $q=12$.}%
\label{Fig:circles - 0<mu<1}%
\end{figure}

\begin{figure}[!htb]
    \centering%
        \subfigure[{$|\xi|\ll r_b$}\label{Fig:roots curve - 0<mu<1 - a}]%
        {\fbox{\includegraphics*[height=2.75cm]{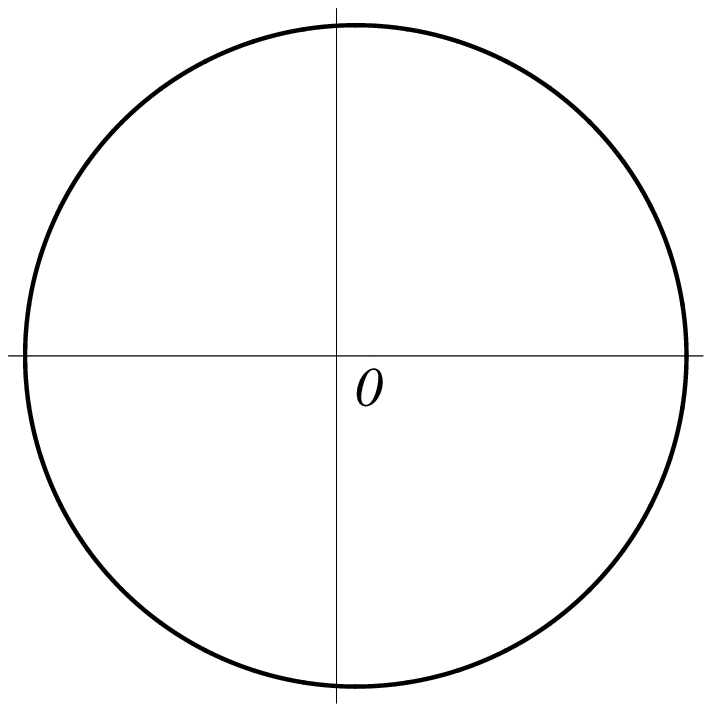}}}\qquad%
        \subfigure[{$|\xi|<r_b$}\label{Fig:roots curve - 0<mu<1 - b}]%
        {\fbox{\includegraphics*[height=2.75cm]{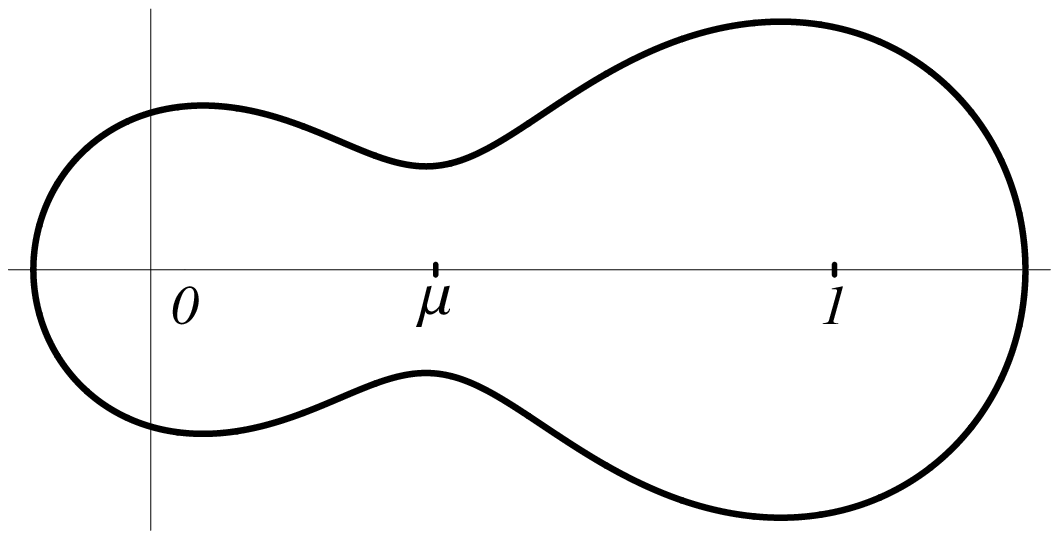}}}\\%
        \subfigure[{$|\xi|=r_b$}\label{Fig:roots curve - 0<mu<1 - c}]%
        {\fbox{\includegraphics*[height=2.75cm]{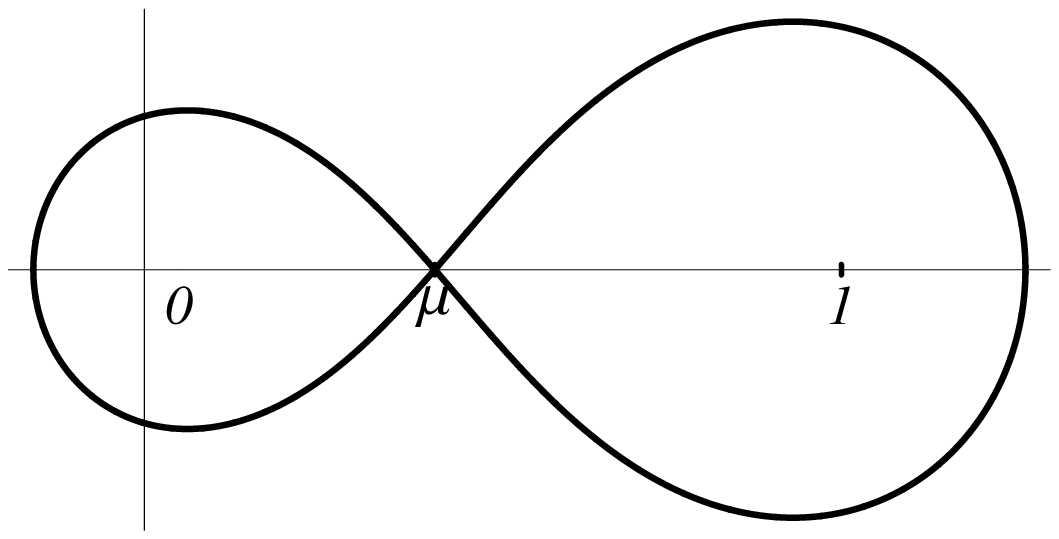}}}\qquad%
        \subfigure[{$|\xi|>r_b$}\label{Fig:roots curve - 0<mu<1 - d}]%
        {\fbox{\includegraphics*[height=2.75cm]{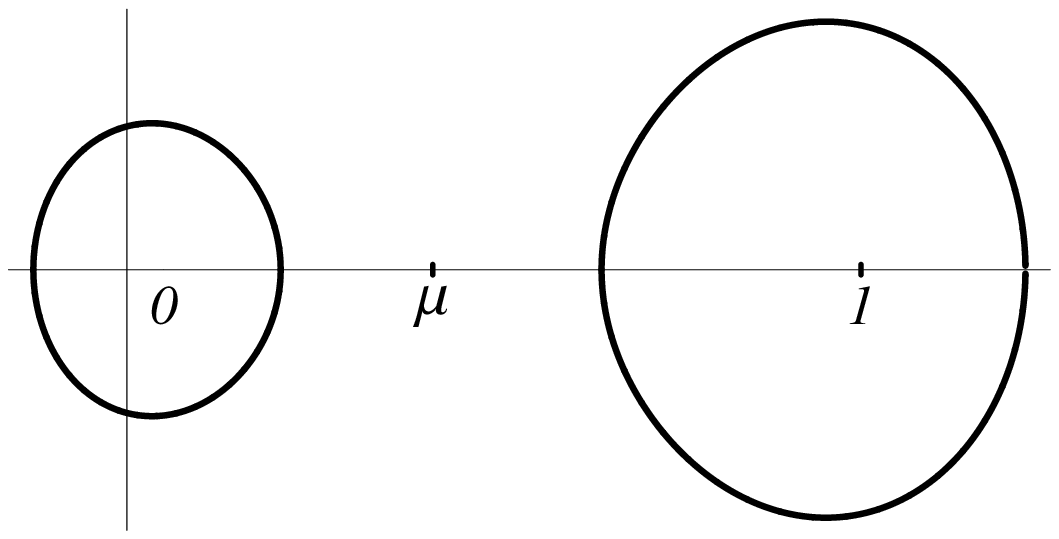}}}%
        \caption{The locus of the roots (\ref{roots curve - 0<mu<1}) on the $w$-plane for different values of $|\xi|$ when $p=5$ and $q=12$.\label{Fig:roots curve - 0<mu<1}}%
\end{figure}

To proceed in labeling the physical root, we rephrase
(\ref{algebraic curve 0<p<q}) in the following form
\begin{eqnarray}\label{algebraic curve 0<p<q:real and imaginary}
\left[(x-1)^2+y^2\right]^{\frac{q-p}{2}}\,\left(x^2+y^2\right)^{\frac{p}{2}}\,|\xi|^q\,e^{i\,\left[q\,\arg{(\xi)}+(q-p)\,\arg{(x-1+i\,y)}+p\,\arg{(x+i\,y)}\right]}=1\,,%
\end{eqnarray}

\noindent where $x=\Re{(w)}$ and $y=\Im{(w)}$ are respectively the
real and imaginary part of $w$ and $\arg{(\xi)}$ is the argument of
the complex variable $\xi$. Equation (\ref{algebraic curve
0<p<q:real and imaginary}) implies that, for fixed $\xi$ and for
rational values of $\mu$, all the roots of (\ref{algebraic curve
0<p<q}) lie on the algebraic curve on the $w$-plane given in
implicit form by
\begin{equation}\label{roots curve - 0<mu<1}
\left[(x-1)^2+y^2\right]^{q-p}\,\left(x^2+y^2\right)^{p}=|\xi|^{-2\,q}\,\,.
\end{equation}
\noindent As implied by the discussion in Section
\ref{Sec:mu-0<mu<1-rational}, for $|\xi|<r_b$ the curve (\ref{roots
curve - 0<mu<1}) consists of a single closed branch that becomes a
big circle of radius $o(|\xi|^{-1})$ around the origin of the
$w$-plane as $|\xi|\rightarrow 0$; for $|\xi|\geq r_b$ the curve
(\ref{roots curve - 0<mu<1}) consists of two closed branches that
become two small circles, one of radius $O(|\xi|^{-\frac{q}{p}})$
around the origin and the other of radius
$O(|\xi|^{-\frac{q}{q-p}})$ around $1$, as $|\xi|\rightarrow \infty$
(see Figure \ref{Fig:roots curve - 0<mu<1}).

\noindent If the independent variable $\xi$ moves counterclockwise
on a circle around the origin, namely if only the phase of $\xi$
varies, then the roots of (\ref{algebraic curve 0<p<q}) move
clockwise on the curve (\ref{roots curve - 0<mu<1}), undergoing
cyclic permutations: if (\ref{roots curve - 0<mu<1}) is constituted
by two unconnected branches, then the two systems of roots undergo
two separate cyclic permutations.

\begin{figure}[!htb]
    \centering%
        \subfigure[{Sector labeling of the $\xi$-plane.}\label{Fig:XiLabeledSectors - 0<mu<1}]%
        {\fbox{\includegraphics*[height=5.8cm]{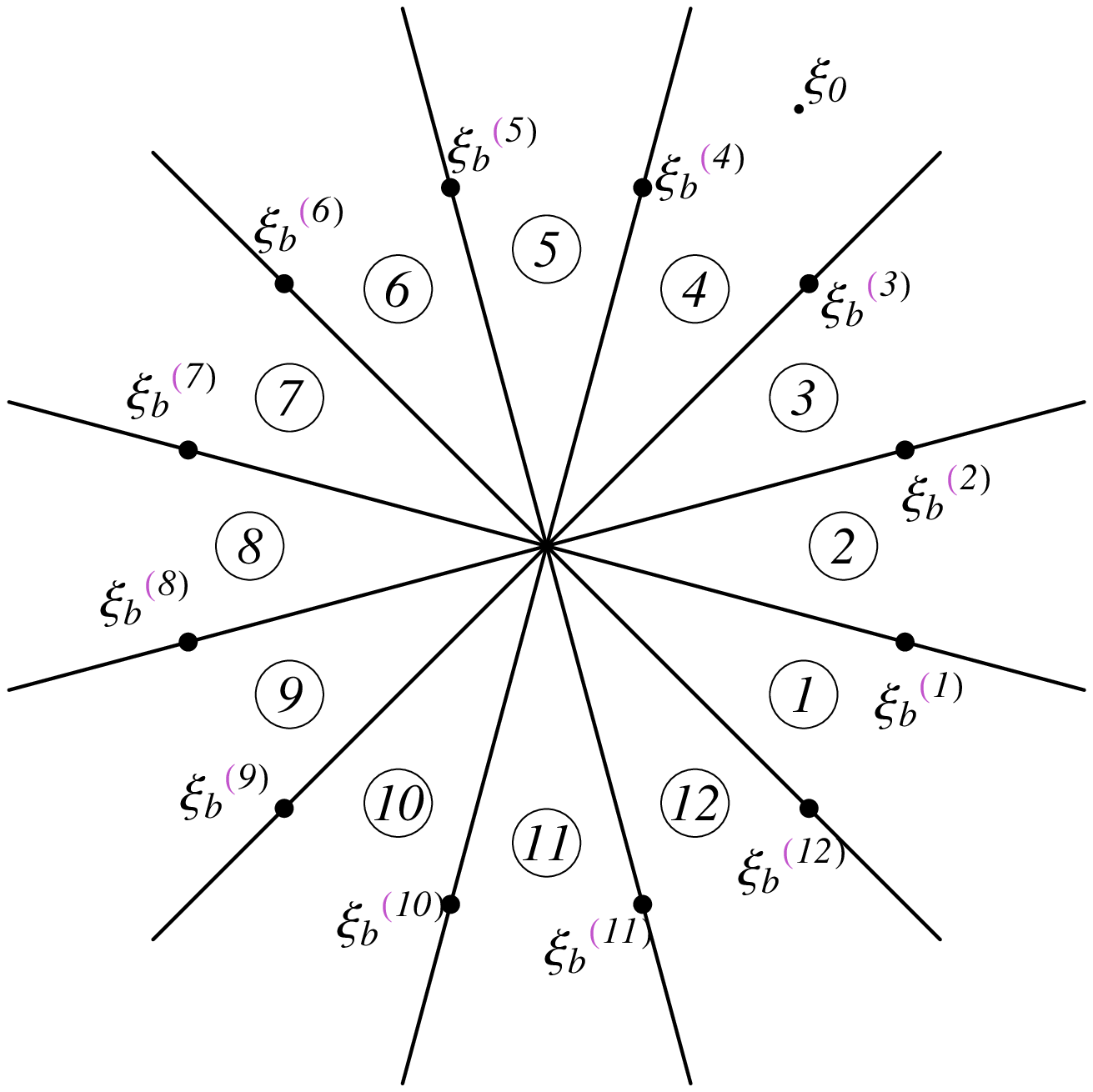}}}\qquad%
        \subfigure[{Corresponding sector labeling of the $w$-plane.}\label{Fig:WLabeledSectors - 0<mu<1}]%
        {\fbox{\includegraphics*[height=5.8cm]{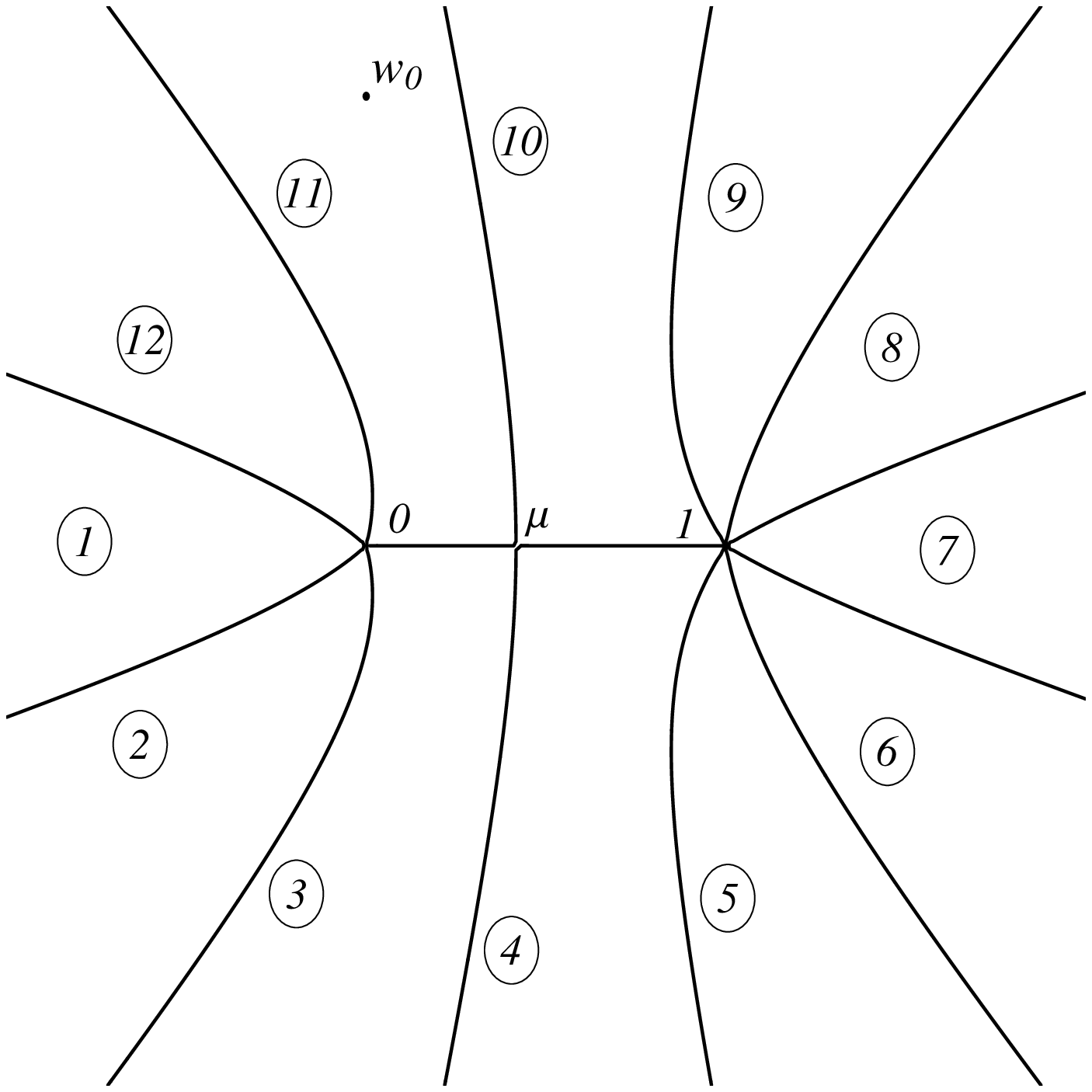}}}%
        \caption{The enumeration scheme for the roots of (\ref{algebraic curve 0<p<q}). In this example $p=5$ and $q=12$, $\xi_0$ is in the fourth sector on the $\xi$-plane, while the physical root $w_0$ lies in the eleventh sector on the $w$-plane.\label{Fig:Labelling - 0<mu<1}}%
\end{figure}

On the other hand, (\ref{SRBP2 0<mu<1}) implies that, if $\xi$ moves
along one of the rays that start at the origin of the $\xi$-plane
and intersect the branch points, \textit{i.e.} if
$\arg{(\xi)}=\arg{(\xi^{(j)}_\mathrm{b})}$ for some $j$, then
$\xi^q\in\mathbb{R}$. Therefore, via (\ref{algebraic curve
0<p<q:real and imaginary}), one immediately gets that in this case
all the roots of (\ref{algebraic curve 0<p<q}) lie on the $w$-plane
on the \textit{locus} defined in implicit form by
\begin{eqnarray}\label{argument curve - 0<mu<1}
(q-p)\,\arg{(x-1+i\,y)}+p\,\arg{(x+i\,y)}=n\,\pi\,\,,\nonumber\\
\mbox{with}\,\left\{\begin{array}{l}n=1,3,5,...,2\,q-1\,\qquad \mbox{if}\,\,(q-p)\,\,\mbox{is odd}\\n=0,2,4,...,2\,q-2\,\qquad\mbox{if}\,\,(q-p)\,\,\mbox{is even .}\end{array}\right.%
\end{eqnarray}
\noindent The branches of the curve (\ref{argument curve - 0<mu<1})
define a partition of the $w$-plane (see Figure \ref{Fig:wSectors -
0<mu<1}). In other words, the partition of the $\xi$-plane into the
$q$ sectors obtained by tracing the rays that start at the origin
and cross the square-root branch points (see Figure
\ref{Fig:XiLabeledSectors - 0<mu<1}) induces a partition of the
$w$-plane into the $q$ sectors individuated by (\ref{argument curve
- 0<mu<1}): if the variable $\xi$ moves entirely in a single angular
sector on the $\xi$-plane, without crossing the lines that define
the sectors, then each root of (\ref{algebraic curve 0<p<q}) is
contained in a single region of the $w$-plane, without crossing any
of the branches of the curve (\ref{argument curve - 0<mu<1}) that
separate the regions.

\begin{figure}[h]
\centering \fbox{\includegraphics[width=8cm]{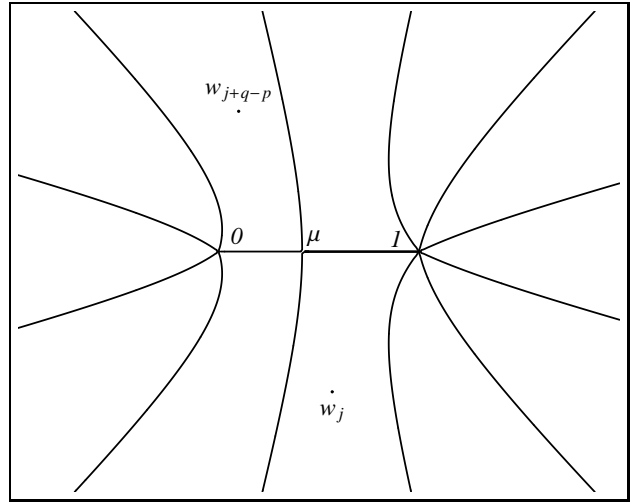}}
\caption{The curve (\ref{argument curve - 0<mu<1}) divides the $w$-plane in $q$ sectors. For a fixed value of $\xi$, each root of (\ref{algebraic curve 0<p<q}) lies inside a sector. When $\xi$ moves counterclockwise around the cut $\gamma_j$, the roots $w_j$ and $w_{q-p+j}$ (that are respectively in the \textit{lower} and \textit{upper exchange region}) exchange their positions in the corresponding sectors. In this example $p=5$ and $q=12$.}%
\label{Fig:wSectors - 0<mu<1}%
\end{figure}

For $q\geq 3$ there is always a branch of the curve (\ref{argument
curve - 0<mu<1}) that intersects the point $\mu$ on the $w$-plane.
As shown in the previous section, when $\xi$ moves around the branch
cut $\gamma_j$ then the two roots $w_j$ and $w_{j+q-p}$ exchange
their positions moving around the point $\mu$ on the $w$-plane (see
Figure \ref{Fig:wSectors - 0<mu<1}). Therefore, if the root $w_j$ is
contained in the lower half-plane of the $w$-plane in the region
that is bounded by the real segment $(\mu,1)$ and by the lower
branch of the curve (\ref{argument curve - 0<mu<1}) that crosses the
point $\mu$ (the \textit{lower exchange region}), then the root
$w_{j+q-p}$ is contained in the upper half-plane of the $w$-plane in
the region that is bounded by the real segment $(0,\mu)$ and by the
upper branch of the curve (\ref{argument curve - 0<mu<1}) that
crosses the point $\mu$ (the \textit{upper exchange region}).

\noindent We assign to each angular sector on the $\xi$-plane the
same label of the branch point that is next to the sector
(counterclockwise), see Figure \ref{Fig:XiLabeledSectors - 0<mu<1}.
If the initial datum $\xi_0$ is contained in the sector $j$ on the
$\xi$-plane, then correspondingly the root $w_{j}$ is in the lower
exchange region on the $w$-plane. Once a label has been assigned to
the lower exchange region, we can label all the remaining regions
(and correspondingly the roots in them) by counting them
counterclockwise $\tmodulus{(q)}$, see Figure
\ref{Fig:WLabeledSectors - 0<mu<1}. When all the regions on the
$w$-plane have been labeled, then the label of the physical root is
the label of the region where $w_{0}$ falls.

The prescription to label the physical root can be summarized as
follows (see Figure \ref{Fig:Labelling - 0<mu<1}):

\begin{enumerate}
\item Trace the circle $B$ in the $\xi$-plane, on which the branch points lie (see \eqref{bpscircle}).
\item Choose arbitrarily the same determination for $\xi_{0}$ and $R$ (see \eqref{InitData-ksi}).
\item Trace the circle $\Xi$ (see \eqref{ksi}) and select, as explained in the present section,
the first branch point $\xi^{(1)}_\mathrm{b}$. Starting from there,
label the remaining branch points counterclockwise. Define the
angular sectors on the $\xi$-plane.
\item Compute $\xi_{0}$ and put it on the
$\xi$-plane. Denote by $\hat j$ the sector where $\xi_{0}$ falls.
\item Trace the curve (\ref{argument curve - 0<mu<1}) on the
$w$-plane. Enumerate the regions defined by the branches of the
curve by assigning $\tmodulus{(q)}$ the label $\hat j$ to the lower exchange region
and moving counterclockwise.
\item From the initial data, compute $w_{0}$ and put it on the
$w$-plane. The label of the physical root is the label of the region
where $w_{0}$ falls.
\end{enumerate}

This mechanism to identify the physical zero cannot be applied when
$\mu$ is irrational as the separatrices of the sectors become
infinitely close.  Therefore this case requires a separate
treatment.

\section{Connection with graph theory}\label{Sec:mu-0<mu<1-rational:Graph Theory}%

In this section we apply graph theory in order to describe the
monodromy group associated to (\ref{algebraic curve 0<p<q}), namely
the subgroup of the symmetrical group $\mathcal{S}_{q}$
corresponding to all possible exchanges and cyclic permutations
occurring on the system of $q$ roots when the independent variable
$\xi$ moves along a closed path on the $\xi$-plane. Since eventually
$\xi$ shall travel along the circle $\Xi$ described by (\ref{ksi})
and the square-root branch points lie on the circle $B$, see (\ref{B
circle radius 0<mu<1 mu rational}), we only need to consider the
inclusions of (one or more) consecutive square-root branch points on
$B$.

We associate a labeled planar graph $V$ to each mutual configuration
of the two circles $\Xi$ and $B$. In order to construct this graph,
let us trace $q$ nodes -- corresponding to the sheets of the Riemann
surface $\Gamma$ -- on the vertices of a regular polygon; let us
arbitrarily give to one of the nodes the label $1$; if $v_{j}$ is
the label of the $j$-th node of the graph, counting counterclockwise
from the node $1$, then the labels are given by the following rule:
$v_1=1$ and $v_{j+1}=(v_{j}+q-p) \tmodulus{(q)}$, $j=1,...,q-1$.
When $\Xi$ includes no square-root branch points, then the planar
graph has no edges (see Figure \ref{Fig:Branch points inclusions p5
q12-a}). If the circle $\Xi$ includes the square-root branch point
$\xi^{(j)}_\mathrm{b}$, $j=1,...,q$, correspondingly the planar
graph gets an edge between the nodes $j$ and $(j+q-p)
\tmodulus{(q)}$, as illustrated in Figure \ref{Fig:Branch points
inclusions p5 q12}. With this prescription, $V$ depends on $\Xi$,
$p$ and $q$, namely $V\equiv V(\Xi,p,q)$.

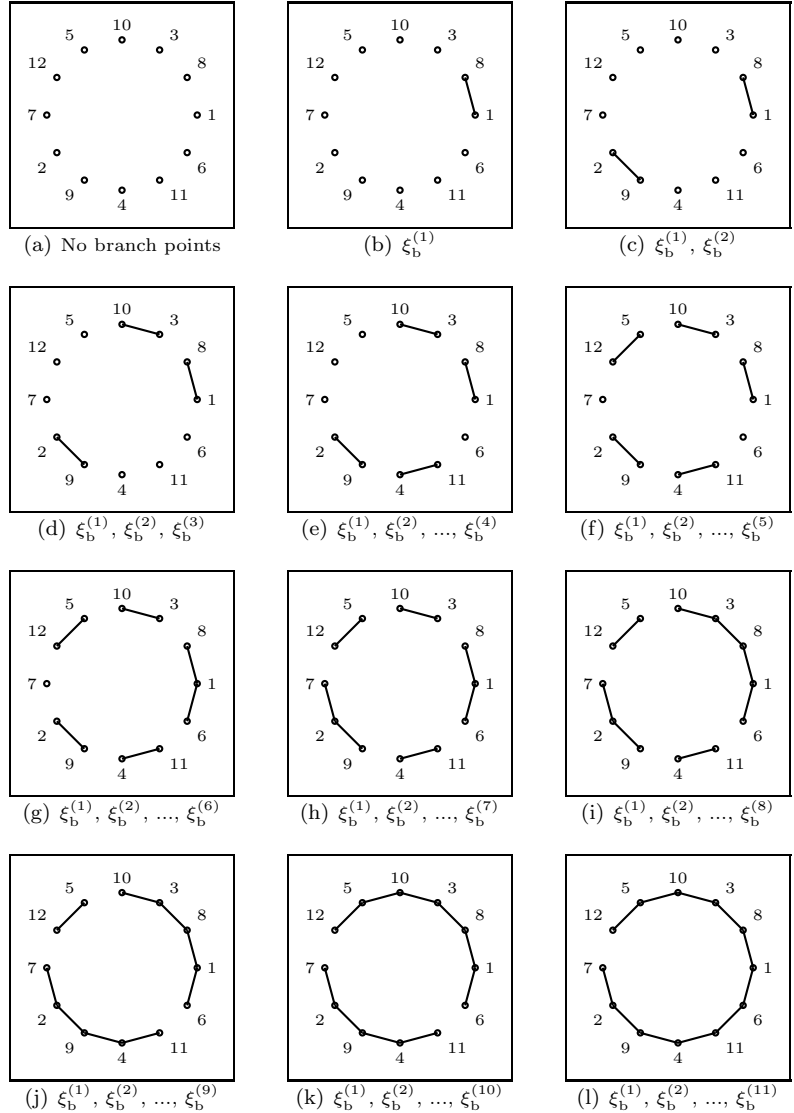
\begin{figure}[h]
\centering
\subfigure[{\scriptsize{No branch points}}\label{Fig:Branch points inclusions p5 q12-a}]{\fbox{\begin{tikzpicture}[style=thick]\dodecagon;\end{tikzpicture}}}\qquad%
\subfigure[{\scriptsize{$\xi^{(1)}_{\mathrm{b}}$}}\label{Fig:Branch points inclusions p5 q12-b}]{\fbox{\begin{tikzpicture}[style=thick]\dodecagon;\draw \VA -- \VB;\end{tikzpicture}}}\qquad%
\subfigure[{\scriptsize{$\xi^{(1)}_{\mathrm{b}}$, $\xi^{(2)}_{\mathrm{b}}$}}\label{Fig:Branch points inclusions p5 q12-c}]{\fbox{\begin{tikzpicture}[style=thick]\dodecagon;\draw \VA -- \VB;\draw \VH -- \VI;\end{tikzpicture}}}\\%
\subfigure[{\scriptsize{$\xi^{(1)}_{\mathrm{b}}$, $\xi^{(2)}_{\mathrm{b}}$, $\xi^{(3)}_{\mathrm{b}}$}}\label{Fig:Branch points inclusions p5 q12-d}]{\fbox{\begin{tikzpicture}[style=thick]\dodecagon;\draw \VA -- \VB;\draw \VH -- \VI;\draw \VC -- \VD;\end{tikzpicture}}}\qquad%
\subfigure[{\scriptsize{$\xi^{(1)}_{\mathrm{b}}$, $\xi^{(2)}_{\mathrm{b}}$, ..., $\xi^{(4)}_{\mathrm{b}}$}}\label{Fig:Branch points inclusions p5 q12-e}]{\fbox{\begin{tikzpicture}[style=thick]\dodecagon;\draw \VA -- \VB;\draw \VH -- \VI;\draw \VC -- \VD;\draw \VJ -- \VK;\end{tikzpicture}}}\qquad%
\subfigure[{\scriptsize{$\xi^{(1)}_{\mathrm{b}}$, $\xi^{(2)}_{\mathrm{b}}$, ..., $\xi^{(5)}_{\mathrm{b}}$}}\label{Fig:Branch points inclusions p5 q12-f}]{\fbox{\begin{tikzpicture}[style=thick]\dodecagon;\draw \VA -- \VB;\draw \VH -- \VI;\draw \VC -- \VD;\draw \VJ -- \VK;\draw \VE -- \VF;\end{tikzpicture}}}\\%
\subfigure[{\scriptsize{$\xi^{(1)}_{\mathrm{b}}$, $\xi^{(2)}_{\mathrm{b}}$, ..., $\xi^{(6)}_{\mathrm{b}}$}}\label{Fig:Branch points inclusions p5 q12-g}]{\fbox{\begin{tikzpicture}[style=thick]\dodecagon;\draw \VA -- \VB;\draw \VH -- \VI;\draw \VC -- \VD;\draw \VJ -- \VK;\draw \VE -- \VF;\draw \VL -- \VA;\end{tikzpicture}}}\qquad%
\subfigure[{\scriptsize{$\xi^{(1)}_{\mathrm{b}}$, $\xi^{(2)}_{\mathrm{b}}$, ..., $\xi^{(7)}_{\mathrm{b}}$}}\label{Fig:Branch points inclusions p5 q12-h}]{\fbox{\begin{tikzpicture}[style=thick]\dodecagon;\draw \VA -- \VB;\draw \VH -- \VI;\draw \VC -- \VD;\draw \VJ -- \VK;\draw \VE -- \VF;\draw \VL -- \VA;\draw \VG -- \VH;\end{tikzpicture}}}\qquad%
\subfigure[{\scriptsize{$\xi^{(1)}_{\mathrm{b}}$, $\xi^{(2)}_{\mathrm{b}}$, ..., $\xi^{(8)}_{\mathrm{b}}$}}\label{Fig:Branch points inclusions p5 q12-i}]{\fbox{\begin{tikzpicture}[style=thick]\dodecagon;\draw \VA -- \VB;\draw \VH -- \VI;\draw \VC -- \VD;\draw \VJ -- \VK;\draw \VE -- \VF;\draw \VL -- \VA;\draw \VG -- \VH;\draw \VB -- \VC;\end{tikzpicture}}}\\%
\subfigure[{\scriptsize{$\xi^{(1)}_{\mathrm{b}}$, $\xi^{(2)}_{\mathrm{b}}$, ..., $\xi^{(9)}_{\mathrm{b}}$}}\label{Fig:Branch points inclusions p5 q12-j}]{\fbox{\begin{tikzpicture}[style=thick]\dodecagon;\draw \VA -- \VB;\draw \VH -- \VI;\draw \VC -- \VD;\draw \VJ -- \VK;\draw \VE -- \VF;\draw \VL -- \VA;\draw \VG -- \VH;\draw \VB -- \VC;\draw \VI -- \VJ;\end{tikzpicture}}}\qquad%
\subfigure[{\scriptsize{$\xi^{(1)}_{\mathrm{b}}$, $\xi^{(2)}_{\mathrm{b}}$, ..., $\xi^{(10)}_{\mathrm{b}}$}}\label{Fig:Branch points inclusions p5 q12-k}]{\fbox{\begin{tikzpicture}[style=thick]\dodecagon;\draw \VA -- \VB;\draw \VH -- \VI;\draw \VC -- \VD;\draw \VJ -- \VK;\draw \VE -- \VF;\draw \VL -- \VA;\draw \VG -- \VH;\draw \VB -- \VC;\draw \VI -- \VJ;\draw \VD -- \VE;\end{tikzpicture}}}\qquad%
\subfigure[{\scriptsize{$\xi^{(1)}_{\mathrm{b}}$, $\xi^{(2)}_{\mathrm{b}}$, ..., $\xi^{(11)}_{\mathrm{b}}$}}\label{Fig:Branch points inclusions p5 q12-l}]{\fbox{\begin{tikzpicture}[style=thick]\dodecagon;\draw \VA -- \VB;\draw \VH -- \VI;\draw \VC -- \VD;\draw \VJ -- \VK;\draw \VE -- \VF;\draw \VL -- \VA;\draw \VG -- \VH;\draw \VB -- \VC;\draw \VI -- \VJ;\draw \VD -- \VE;\draw \VK -- \VL;\end{tikzpicture}}}\\%
\caption{Planar graphs associated to different inclusions of consecutive square-root branch points in the circle $\Xi$, for $p=5$ and $q=12$. In each caption, the included branch points are indicated.}\label{Fig:Branch points inclusions p5 q12}%
\end{figure}

From the general theory we know that the inclusion of exactly $q$
square-root branch points inside the circle $\Xi$ is equivalent to
the inclusion of the branch point at infinity: in this case the
period of the zeros of (\ref{algebraic curve 0<p<q}) are $q-p$ or
$p$, see (\ref{Ansatz1a})-(\ref{Ansatz1c}) respectively
(\ref{Ansatz2a})-(\ref{Ansatz2c}), with the first $q-p$ roots having
period $q-p$ and the last $p$ roots having period $p$. From now on,
we will suppose to have at most $q-1$ consecutive square-root branch
points, starting from $\xi^{(1)}_{\mathrm{b}}$, inside the circle
$\Xi$.

A \textit{path} of length $n$ on the graph $V$ is a sequence of $n$
adjacent distinct nodes $\{v_{j_1},v_{j_2},...,v_{j_n}\}$, namely
such that $(v_{j_1},v_{j_2})$, $(v_{j_2},v_{j_3})$, ...,
$(v_{j_{n-1}},v_{j_n})$ are edges of the graph. For fixed $\Xi$, $p$
and $q$, if in $V(\Xi,p,q)$ the node $v_j$ belongs to a path of
length $b$, then the period of the corresponding root of the
algebraic equation (\ref{algebraic curve 0<p<q}), $w_{v_{j}}(\xi)$,
is exactly $b$, while $\xi$ moves along the circle $\Xi$. If the
node $v_j$ is unconnected (\textit{i.e.} it is touched by no edges),
then the period of $w_{v_{j}}(\xi)$ is $1$. For fixed $p$ and $q$
and for a certain mutual configuration of the two circles $\Xi$ and
$B$, in order to describe the behavior of the roots of
(\ref{algebraic curve 0<p<q}), we need to measure the lengths of the
paths in the corresponding graph $V(\Xi,p,q)$.

For a graph $V(\Xi,p,q)$ with $q$ nodes, the set of all path lengths
corresponds to a decomposition in cycles of elements of the
symmetrical group $\mathcal{S}_{q}$.
Once the correspondence between the graphs $V$'s and the elements of
$\mathcal{S}_{q}$ has been established, we can use the well-known
correspondence between Ferrer diagrams and permutations. We recall
that a Ferrer diagram is an ordered disposition of $q$ blank boxes
in rows and columns, such that the number of boxes in each column
equals the lengths of the cycles of the corresponding permutation
and the total number of columns is the number of cycles in which the
permutation is decomposed. For instance, the Ferrer diagram
\begin{center}
\[\yng(6,5,1)\]
\end{center}
\noindent corresponds to the planar graph presented in Figure
\ref{Fig:Branch points inclusions p5 q12-g}. In a Ferrer diagram,
the columns are sorted in decreasing order of length, from left to
right. Moreover, note that each column in a Ferrer diagram is
oriented up-down, the first row being at the top, the last row being
at the bottom. Then the problem of knowing the period of a root of
equation (\ref{algebraic curve 0<p<q}), while $\xi$ moves along the
circle $\Xi$, is reduced to measuring the lengths of the columns of
an appropriate Ferrer diagram.

\subsection{The bumping rule}\label{Sec:mu-0<mu<1-rational:The bumping rule}%

In this subsection we show how to recursively build the Ferrer
diagrams associated to the planar graphs $V$'s. For fixed $p$ and
$q$, suppose that the circle $\Xi$ includes $b$ square-root branch
points, so that $V(\Xi,p,q)$ is the corresponding labeled planar
graph. In the following, we include only consecutive square-root
branch points of $B$ in $\Xi$, always starting from
$\xi^{(1)}_{\mathrm{b}}$ and moving counterclockwise:
$\{\xi^{(1)}_{\mathrm{b}},\xi^{(2)}_{\mathrm{b}},...,\xi^{(b)}_{\mathrm{b}}\}$.
Under this hypothesis, $V(\Xi,p,q)$ depends on $p$ and $q$ and on
$\Xi$ only via the number $b$ of included square-root branch points.
Assuming that $p$ and $q$ have been fixed once for all, we set
$V(\Xi,p,q)\equiv V(b)$. Let $F(b)$ indicate the Ferrer diagram
corresponding to the planar graph $V(b)$. In what follows, we
construct a recursive \textit{bumping rule} for Ferrer diagrams,
namely a rule that permits to build $F(b+1)$ from $F(b)$.

First of all, we need two integer sequences:
\begin{subequations}
\begin{eqnarray} \label{q and p sequences}
  q_{k} = p_{k-1} \,,\qquad  &&q_{0}=q\,,\\
  p_{k} = q_{k-1} \modulus{(p_{k-1}}) \,,\qquad &&p_{0}=q-p \,,
\end{eqnarray}
\noindent with $0 \leq k \leq \bar{k}$, where $\bar{k}$ is an
integer number such that
\begin{equation}\label{k barra}
    q_{\bar{k}}=1 \mbox{ ;}
\end{equation}
\end{subequations}

\noindent note that $\bar{k}$ always exists due to the decreasing
nature of the sequence $\{q_{k}\}$. We need also an auxiliary
recursive sequence, written as a combination of the previous two,
\begin{subequations}
\begin{equation}\label{b sequence}
    b_{k}=b_{k-1}+q_{k-1}-p_{k-1} \,\,\mbox{ , }\,\,b_{0}=0
    \,\,\mbox{ .}
\end{equation}
\noindent We remark here that the $b$-sequence (\ref{b sequence})
divides the discrete segment $[1,q-1]$ into $\bar{k}$ parts of
length $q_{k}-p_{k}$. Moreover, from the first of the (\ref{q and p
sequences}), via (\ref{b sequence}), one gets
\begin{eqnarray}\label{b sequence in terms of q}
b_{0}=0 \,\, , \,\, b_{k} &=& b_{k-1}+q_{k-1}-q_{k}=b_{k-2}+q_{k-2}-q_{k-1}+q_{k-1}-q_{k} \nonumber \\
                          &=& b_{k-2}+q_{k-2}-q_{k}=...=b_{k-h}+q_{k-h}-q_{k} \nonumber \\
                          &=& b_{0}+q_{0}-q_{k}= q-q_{k} \mbox{ .}
\end{eqnarray}
\end{subequations}

Suppose that $\Xi$ includes $b$ square-root branch points. If we
modify $\Xi$ so that it includes also the next adjacent branch
point, correspondingly we modify $V(b)$ into $V(b+1)$, adding to
$V(b)$ a new edge between the nodes $(b+1)$ and
$(b+1+q-p)\tmodulus{(q)}$.

If $0\leq b<b_{1}-1=q-p-1$, then each new edge added to the graph
$V(b)$ connects a single unconnected node to a path composed by a
certain number of edges (at least, to another single unconnected
node). When $b=b_{1}-1$, it is impossible to trace a new edge on
$V(b)$ connecting two unconnected nodes or a single node to a path:
the insertion of a new edge in a graph at this point causes the
connection of two paths, resulting in a new longer path in the graph
$V(b+1)$.

To understand what happens when $b_{1}\leq b<b_{2}-1$, let us
imagine to build an auxiliary graph in the following way: associate
a weighted planar graph $\tilde{V}(b)$ to the graph $V(b)$, so that,
for each path on $V(b)$, you have a weighted node in $\tilde{V}(b)$,
with weights on $\tilde{V}(b)$ equal to the lengths of the paths on
$V(b)$ (we recall that a single node is equivalent to a path of
length $1$). Now the discussion of the case $b_{1}\leq b<b_{2}-1$ is
analogous to the discussion of the previous case $0\leq b<b_{1}-1$,
using $\tilde{V}(b)$ instead of $V(b)$, with the only difference
that the length of a path in $\tilde{V}(b)$ is not just the number
of nodes touched by the path, but the sum of all the weights of the
nodes touched by the path. If $b_{1}\leq b<b_{2}-1$, then each new
edge added to the auxiliary graph $\tilde{V}(b)$ connects a single
unconnected weighted node to a weighted path composed by a certain
number of edges (at least, to another single unconnected weighted
node). But when $b=b_{2}-1$, again it is impossible to trace a new
edge connecting two unconnected weighted nodes or a single weighted
node to a weighted path on $\tilde{V}(b)$: the insertion of a new
edge in $\tilde{V}(b)$ at this point will cause the connection of
two weighted paths, producing a new longer path in the auxiliary
graph.

In order to progressively include more branch points inside $\Xi$,
we must build each time a new auxiliary graph, in the
afore-described way. We iterate this operation, building a new
auxiliary weighted graph every time we arrive to include exactly
$b_{k}$ branch points inside the circle $\Xi$, until we reach
$b=q-1$. In this language, $q_{k}$ is the number of unconnected
weighted nodes in each auxiliary graph when $b_{k} \leq b <
b_{k+1}$, while $p_{k}$ is the number of weighted nodes in each
auxiliary graph when exactly $b_{k+1}$ branch points are included in
$\Xi$.

The use of Ferrer diagrams strongly simplifies this picture. When
$b=b_{k}$, the lengths of the columns of the Ferrer diagram $F(b)$
represent the values of the weights of the nodes of the
corresponding auxiliary weighted graph. If $b_{k} < b < b_{k+1}$,
then the lengths of the columns of the corresponding Ferrer diagrams
$F(b)$ represent the lengths of the weighted paths in the
corresponding auxiliary weighted graphs. When $b=b_{k}$, the number
of columns in $F(b)$ is $q_{k}$. Moreover, note that, if one starts
to count columns from the left in $F(b)$, when $b_{h} \le b <
b_{h+1}$, then the columns lying at the positions $p_{h}+1$,
$p_{h}+2$,..., have all the same length, since they coincide with
the shortest paths on the corresponding planar graphs.

If the circle $\Xi$ includes $b$ consecutive square-root branch
points, when $b_{j} \le b < b_{j+1}$, and if we modify it in order
to include $b+1$ consecutive square-root branch points, then we can
build $F(b+1)$ just moving the whole last column on the right of the
Ferrer diagram $F(b)$ under the first available column from the
left, never occupying positions beyond the $p_{j}$-th column. To
pass from $b_{j}$ to $b_{j+1}$, see (\ref{b sequence}), we have to
move $q_{j}-p_{j}$ columns.
All the above considerations lead to the following prescription.

\noindent \textbf{The bumping rule.} \textit{The Ferrer diagram
$F(0)$, corresponding to the inclusion of zero branch points, is
composed by a single row of $q$ columns of length $1$. The Ferrer
diagram $F(b)$, with $b\leq q-1$, can be obtained from the Ferrer
diagram $F(b-1)$ in the following way:
\begin{itemize}
\item compute $h$ such that $b_{h} \le b < b_{h+1}$;%
\item counting the columns of $F(b-1)$ from the left, move its whole
column located at the position $p_{h}+1$ under the column located at
the position $(b-b_{h})\tmodulus{(p_h)}$.
\end{itemize}
}
\noindent We remark that, in principle, one could ask to move any
column lying at the position $p_{h}+1$, $p_{h}+2$, ..., since they
all have the same length. The reason for prescribing a movement of
the column lying at the position $p_{h}+1$ will be clear after the
introduction in Subsection \ref{Sec:mu-0<mu<1-rational:Period
Formula} of a numeration for the blank boxes of the Ferrer diagrams.

\begin{figure}[h]
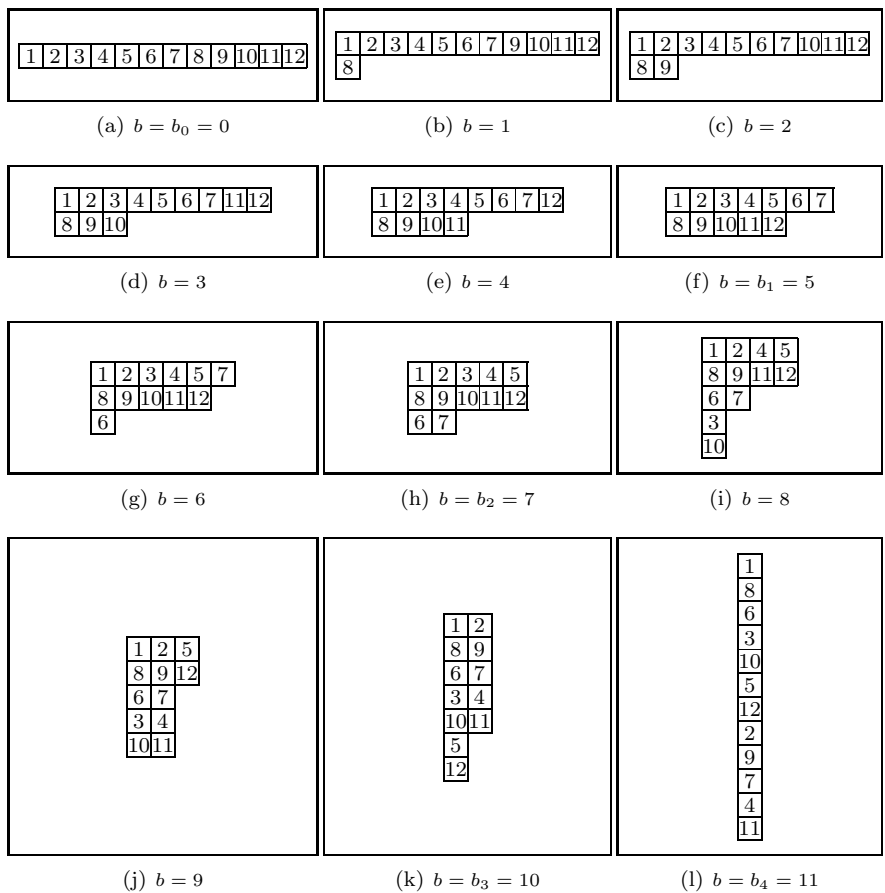

\setlength{\unitlength}{0.1cm} \centering%
\subfigure[{\scriptsize{$b=b_0=0$}}\label{Fig:Ferrer diagrams p5 q12-a}]{\framebox(41,12){\footnotesize{\young(123456789\ten\eleven\twelve)}}}\hspace{0.1cm}%
\subfigure[{\scriptsize{$b=1$}}\label{Fig:Ferrer diagrams p5 q12-b}]{\framebox(38,12){\footnotesize{\young(12345679\ten\eleven\twelve,8)}}}\hspace{0.1cm}%
\subfigure[{\scriptsize{$b=2$}}\label{Fig:Ferrer diagrams p5 q12-c}]{\framebox(35,12){\footnotesize{\young(1234567\ten\eleven\twelve,89)}}}\\%
\subfigure[{\scriptsize{$b=3$}}\label{Fig:Ferrer diagrams p5 q12-d}]{\framebox(41,12){\footnotesize{\young(1234567\eleven\twelve,89\ten)}}}\hspace{0.1cm}%
\subfigure[{\scriptsize{$b=4$}}\label{Fig:Ferrer diagrams p5 q12-e}]{\framebox(38,12){\footnotesize{\young(1234567\twelve,89\ten\eleven)}}}\hspace{0.1cm}%
\subfigure[{\scriptsize{$b=b_1=5$}}\label{Fig:Ferrer diagrams p5 q12-f}]{\framebox(35,12){\footnotesize{\young(1234567,89\ten\eleven\twelve)}}}\\%
\subfigure[{\scriptsize{$b=6$}}\label{Fig:Ferrer diagrams p5 q12-g}]{\framebox(41,20){\footnotesize{\young(123457,89\ten\eleven\twelve,6)}}}\hspace{0.1cm}%
\subfigure[{\scriptsize{$b=b_2=7$}}\label{Fig:Ferrer diagrams p5 q12-h}]{\framebox(38,20){\footnotesize{\young(12345,89\ten\eleven\twelve,67)}}}\hspace{0.1cm}%
\subfigure[{\scriptsize{$b=8$}}\label{Fig:Ferrer diagrams p5 q12-i}]{\framebox(35,20){\footnotesize{\young(1245,89\eleven\twelve,67,3,\ten)}}}\\%
\subfigure[{\scriptsize{$b=9$}}\label{Fig:Ferrer diagrams p5 q12-j}]{\framebox(41,42){\footnotesize{\young(125,89\twelve,67,34,\ten\eleven)}}}\hspace{0.1cm}%
\subfigure[{\scriptsize{$b=b_3=10$}}\label{Fig:Ferrer diagrams p5 q12-k}]{\framebox(38,42){\footnotesize{\young(12,89,67,34,\ten\eleven,5,\twelve)}}}\hspace{0.1cm}%
\subfigure[{\scriptsize{$b=b_4=11$}}\label{Fig:Ferrer diagrams p5 q12-l}]{\framebox(35,42){\footnotesize{\young(1,8,6,3,\ten,5,\twelve,2,9,7,4,\eleven)}}}\\%
\caption{Numbered Ferrer diagrams associated to different inclusions of consecutive square-root branch points in the circle $\Xi$, for $p=5$ and $q=12$. In each caption, the number of included branch points $b$ is indicated. The numeration is explained in the text.}\label{Fig:Ferrer diagrams p5 q12}%
\end{figure}

For each $k$ between $0$ and $\bar{k}$, (\ref{k barra}), the Ferrer
diagram $F(b_{k})$ features $q_{k}$ columns of two possible lengths
only, $p_{k}$ columns of length $T^{(1)}_{k}$ and $q_{k}-p_{k}$
columns of length $T^{(2)}_{k}$ (see the diagrams in Figure
\ref{Fig:Ferrer diagrams p5 q12}, neglecting the numeration, that
will be explained below). A \textit{$k$-level Ferrer diagram} is a
Ferrer diagram with exactly $q_{k}$ columns.

The bumping rule implies that we ``pass'' from a $k$-level Ferrer
diagram to a $(k+1)$-level Ferrer diagram once we have moved all the
shortest columns on the right under the tallest columns on the left.


The column lengths, $T^{(1)}_{k}$ and $T^{(2)}_{k}$, in a $k$-level
Ferrer diagram can be given by the following recursive rule:
\begin{equation} \label{Ferrer heights}
   \left \{
         \begin{array}{lcc}
         T^{(1)}_{k} p_{k} + T^{(2)}_{k} (q_{k}-p_{k})=q_{0}\,, &  & T^{(1)}_{0}=1\,,\\ \\
         T^{(1)}_{k+1} = T^{(1)}_{k} +
         T^{(2)}_{k} \left( \left \lfloor \frac{q_{k}}{p_{k}} \right \rfloor +1 \right)\,, & & T^{(2)}_{0}=1  \,,
                  \end{array}
   \right.
\end{equation}
\noindent where $\lfloor x \rfloor$ is the floor of the number $x$
(namely, the largest integer less than or equal to $x$). Indeed, the
first of the (\ref{Ferrer heights}) is a sort of conservation rule
for the number of boxes in a $k$-level Ferrer diagram. The second of
the (\ref{Ferrer heights}) comes directly from the bumping rule: at
each level, the length of the first column is the sum of the length
of the column at the previous level plus the length of the last
column multiplied by the number of the necessary column movements.
Although relations (\ref{Ferrer heights}) are enough to determine
the two quantities $T^{(1)}$ and $T^{(2)}$, it is more convenient to
put them in a different form:
\begin{equation} \label{T1 T2}
   \left \{
         \begin{array}{lcc}
         T^{(1)}_{k+1} = T^{(1)}_{k} +
         T^{(2)}_{k} \left( \left \lfloor \frac{q_{k}}{p_{k}} \right \rfloor +1 \right) \,,&  & T^{(1)}_{0}=1\,, \\ \\
         T^{(2)}_{k+1} = T^{(1)}_{k} +
         T^{(2)}_{k} \left( \left\lfloor \frac{q_{k}}{p_{k}} \right\rfloor \right) \,, &  & T^{(2)}_{0}=1\,.
         \end{array}
   \right.
\end{equation}
\noindent In the next subsection, we present a useful link between these
two-steps-recursive relations, and the continued-fraction expansion
of the number $\frac{1}{1-\mu}$.

\subsection{Connection with continued fractions}\label{Sec:mu-0<mu<1-rational:continued fractions}%

Let us recall some basic notions about simple continued fractions.
Let $x$ be a non-negative real number. We associate to the real
number $x$ an integer sequence $\{a_{k}\}$ such that:
\begin{equation}\label{simple continued fractions}
x=a_{0}+\frac{1}{a_{1}+\frac{1}{a_{2}+\frac{1}{a_{3}+...}}} \mbox{
.}
\end{equation}
\noindent We say that $\langle a_{0}, a_{1}, a_{2}, ...\rangle$ is
the simple continued fraction of $x$ with $a_{k}$ positive
integers $\forall\,k>0$ (as usual, for the sake of simplicity, we
avoid repeating the adjective simple when referring to
simple continued fractions). The elements $a_{k}$ of the
continued fraction expansion are called \textit{partial quotients}.

The integer sequence $\{a_{k}\}$ is finite if and only if
the corresponding number $x$ is a rational number. The number of
elements in the sequence is called the \textit{length} of the
continued fraction.


The partial quotients can be recursively calculated introducing the
auxiliary sequence of the \textit{remainders} $r_{k}$:
\begin{equation}\label{generic r sequence}
\left\{
\begin{array}{ll}
    r_{k}=\frac{1}{r_{k-1}-\lfloor r_{k-1}\rfloor}\,,\qquad r_{0}=x, \\
    a_{k}=\lfloor r_{k}\rfloor\end{array}%
\right.
\end{equation}
We trust no reader will be confused by the similarity of this standard notation for the \textit{remainders} $r_{k}$
with the radius $r_{b}$, see \eqref{B circle radius 0<mu<1 mu rational}. The rational number $c_{n}\equiv c_{n}(x)$ obtained truncating the
continued fraction of $x$ at the $n$-th term is called the
\textit{$n$-th convergent} of the continued fraction:
\begin{equation}\label{n-th convergent}
c_{n}=\frac{P_{n}}{Q_{n}}=\langle
a_{0},a_{1},...,a_{n}\rangle=a_{0}+\frac{1}{a_{1}+\frac{1}{...+\frac{1}{a_{n}}}}
\mbox{ .}
\end{equation}
\noindent The numerator $P_{n}$ and the denominator $Q_{n}$ of the
$n$-th convergent $c_{n}$ satisfy the following second-order
recurrence relations:
\begin{equation}\label{convergent numerator and denominator}
    \begin{array}{ccc}
      P_{n}=a_{n} P_{n-1}+P_{n-2}, \qquad & P_{-2}=0, & P_{-1}=1; \\
      Q_{n}=a_{n} Q_{n-1}+Q_{n-2}, \qquad & Q_{-2}=1, & Q_{-1}=0. \\
    \end{array}
\end{equation}

Now we are ready to rephrase the quantities introduced in the
previous subsection in the language of  continued fractions.
First of all, let us eliminate from the $q$-sequence the dependence
on the $p$-sequence in formula (\ref{q and p sequences}):
\begin{equation}\label{q sequence}
      q_{n}=q_{n-2} \modulus{(q_{n-1})}, \qquad  q_{0}=q,\quad  q_{1}=q-p \,. \\
\end{equation}
\noindent We can also write this formula as follows:%
\begin{equation}\label{q sequence fraction}
      \frac{q_{n}}{q_{n-1}}=\frac{q_{n-2}}{q_{n-1}}-\left\lfloor \frac{q_{n-2}}{q_{n-1}} \right\rfloor \,,\qquad  q_{0}=q ,\,\, q_{1}=q-p \mbox{ ,}
\end{equation}
\noindent where we used the fact that $\frac{x-[x
\modulus{(y)}]}{y}=\left\lfloor \frac{x}{y} \right\rfloor$ for  all
$x,y\in\mathbb{N}^{+}$. Now, let us set
\begin{equation}\label{r in terms of q}
    r_{k}=\frac{q_{k}}{q_{k+1}} \,\, \mbox{ with } 0 \leq k \leq \bar{k}\,\,\mbox{ .}
\end{equation}
\noindent Using (\ref{q sequence fraction}), we recover formula
(\ref{generic r sequence}) for the quantity $r_{k}$,
\begin{equation}\label{r sequence}
r_{k}=\frac{1}{r_{k-1}-\lfloor r_{k-1}\rfloor} \,\, \mbox{ with }
\,\, r_{0}=\frac{q_{0}}{q_{1}}=\frac{q}{q-p}=\frac{1}{1-\mu} \,\,
\mbox{ .}
\end{equation}
\noindent The $q$-sequence, namely the number of columns in the
Ferrer diagrams, can be obtained by inverting formula (\ref{r in
terms of q}),
\begin{equation}\label{q in terms of r}
    q_{k}=\frac{q_{k-1}}{r_{k-1}}=q \left(\prod_{j=0}^{k-1}\frac{1}{r_{j}}\right) \,,\qquad q_{0}=q ,\,\, r_{0}=\frac{1}{1-\mu} ,\qquad 0 \leq k \leq \bar{k}\,\,\mbox{ ,}
\end{equation}
\noindent where the $r_{k}$ are the terms of the sequence of the
remainders (see(\ref{r sequence})) of the continued fraction
expansion of the number $\frac{1}{1-\mu}$, with partial quotients
\begin{equation}\label{a sequence}
    a_{k} = \left\lfloor r_{k} \right\rfloor = \left\lfloor \frac{q_{k}}{p_{k}} \right\rfloor
\end{equation}
\noindent and length $\bar{k}$. The $p$-sequence can be calculated
by using the relation:
\begin{equation}\label{p sequence}
    p_{k}=q_{k+1} \,,\qquad  0 \leq k \leq \bar{k}\,.
\end{equation}

We can moreover provide a convenient reformulation yielding the
lengths of the columns in the Ferrer diagrams (\ref{T1 T2}),
$T^{(1)}_{k}$ and $T^{(2)}_{k}$. Indeed, by subtracting the two
recursion relations (\ref{T1 T2}), one gets
\begin{equation}\label{T1 in terms of T2}
    T^{(1)}_{k+1}=T^{(2)}_{k+1}+T^{(2)}_{k} \,,
\end{equation}
\noindent and the insertion of this last formula in the second of
the (\ref{T1 T2}) yields, via (\ref{a sequence}), a $3$-terms linear
recursion for $T^{(2)}_{k}$:
\begin{equation}\label{T2 in terms of T2}
    T^{(2)}_{k+1}=a_{k}\,T^{(2)}_{k}+T^{(2)}_{k-1} \,.
\end{equation}
\noindent A comparison of this last recursion relation with
(\ref{convergent numerator and denominator}), and an analogous
comparison of the starting conditions, yield the relation
\begin{equation}\label{T2}
    T^{(2)}_{k}=P_{k-1} \,,
\end{equation}
\noindent where $P_{k}$ is the numerator of the $k$-th convergent of
the continued fraction expansion of $\frac{1}{1-\mu}$. Then,
(\ref{T1 in terms of T2}) and (\ref{T2}) imply
\begin{equation}\label{T1}
    T^{(1)}_{k}=P_{k-1}+P_{k-2} \,.
\end{equation}

Now all the quantities involved in the description of the Ferrer
diagrams (\textit{i.e.} the numbers of columns and boxes in each
column) are expressed in terms of the continued fraction expansion
of the (rational) number $\frac{1}{1-\mu}$.

We conclude this section by noting a more convenient way of writing
the fundamental $b$-sequence (\ref{b sequence}). Indeed, by
combining (\ref{r in terms of q}), (\ref{r sequence}) and (\ref{a
sequence}), one gets: \begin{subequations}
\begin{equation}\label{q in terms of a}
q_{k}=q_{k-2}-a_{k-2} q_{k-1} \,,\qquad q_{0}=q \,,\quad q_{1}=q-p
\,\,\mbox{ ,}
\end{equation}
\noindent and by comparing this last relation with the recursion
relations (\ref{convergent numerator and denominator}) via the
following \textit{ansatz}:
\begin{equation}\label{q ansatz}
q_{k}=(-1)^{k} \left(\alpha P_{k-2}+\beta Q_{k-2}\right) \,\,,\,\, k
\geq 0 \,\,\mbox{ ,}
\end{equation}
\noindent one immediately gets
\begin{equation}\label{alpha beta}
\alpha=(p-q) \,,\qquad \beta=q \,;
\end{equation}
\end{subequations}
\noindent the insertion of (\ref{q ansatz}) and  (\ref{alpha beta})
in (\ref{b sequence in terms of q}) entails
\begin{equation}\label{b sequence in terms of P and Q}
b_{k}=q-q_{k}=q-(-1)^{k}\left[(p-q)\,P_{k-2}+q\,Q_{k-2}\right]
\,,\qquad k\geq 0 \,.
\end{equation}

\section{Determination of the period}

\subsection{The period formula for $\mu\in\mathbb{Q}$}\label{Sec:mu-0<mu<1-rational:Period Formula}%

Once the structure of the Riemann surface has been achieved, our
next step is to provide an explicit formula to predict the period of
each of the $q$ roots of the algebraic equation (\ref{algebraic
curve 0<p<q}).

The variable $\xi$ moves along the circle $\Xi$ which includes $b$
consecutive square-root branch points on the $\xi$-plane, starting
from $\xi^{(1)}_{\mathrm{b}}$. Now we introduce a numbering of the
boxes of the Ferrer diagrams, so that, if one fixes a single column,
all the roots labeled with the numbers appearing in the chosen
column have the same period, equal to the length of the column. For
instance, look at Figure \ref{Fig:Ferrer diagrams p5 q12-i} when
$b=8$. We interpret this picture, inferring that the roots $w_1$,
$w_3$, $w_6$, $w_8$ and $w_{10}$ have period $5$; the roots $w_2$,
$w_7$ and $w_9$ have period $3$; the roots $w_4$, $w_5$, $w_{11}$
and $w_{12}$ have period $2$. This is exactly what you obtain by
comparing with Figure \ref{Fig:Branch points inclusions p5 q12-i}.

For fixed $p$ and $q$, if we enumerate the boxes of $F(0)$, starting
from the left, with the first consecutive $q$ natural numbers
($1$,$2$,...,$q$) and if we apply recursively the bumping rule, then
we obtain the correct enumeration for each $F(b)$ with $b>0$.

Let us introduce the number $l_{h}(s)$, defined recursively as
follows:
\begin{equation}\label{lsymbol}
l_{h}(s)=l_{h-1}(s) \tmodulus{(p_{h-1})} \,,\qquad
l_{0}(s)=s \,,
\end{equation}

\noindent for all $s=1,2,...,q$ and for all $h=1,2,...,\bar{k}$.
This number $l_{h}(s)$ gives the position of the column where the
box with the number $s$ lies in the $h$-level Ferrer diagram. Once
we know this information for a $h$-level Ferrer diagram, we can
extend the same information for a generic Ferrer diagram.

Let us suppose that there are $b$ branch points included in the
circle $\Xi$, with $b_{h} \leq b < b_{h+1}$. We are interested in
the period of $w_s$. Then two cases are possible.
\begin{description}
\item[Case 1] If $l_{h}(s) > b-b_{h}+p_{h}$ then
    $T^{(2)}_{h}$ is the length of the column where the box with number
    $s$ lies; this column was not moved from the position it
    occupied when exactly $b_{h}$ branch points were included in the
    circle $\Xi$.
\item[Case 2] If $l_{h}(s) \leq b-b_{h}+p_{h}$ then we must
    consider two subcases: if $l_{h}(s) \leq p_{h}$, then the length of
    the column in $F(b_h)$, containing the box $s$ and lying at the position
    $l_{h}(s)$, is $T^{(1)}_{h}$; if $l_{h}(s) > p_{h}$, then the length
    of the column in $F(b_h)$, containing the box $s$ and lying at
    the position $l_{h}(s)$, is $T^{(2)}_{h}$. In both cases, in order to
    obtain $F(b)$ starting from $F(b_h)$, the bumping rule predicts
    that, if the circle $\Xi$ includes other $b-b_h$ branch points, then
    some of the last $(q_{h}-p_{h})$ columns in $F(b_h)$, with length
    $T^{(2)}_{h}$, are moved under the column located at the position
    $l_{h}(s) \tmodulus{(p_{h})}$. If $l_{h}(s) \leq p_{h}$, then
    $l_{h}(s) \tmodulus{(p_{h})}=l_{h}(s)$ and the column at the
    position $l_{h}(s)$ in $F(b)$ contains the box $s$. If $l_{h}(s) > p_{h}$,
    then the column in $F(b_h)$, at the position $l_{h}(s)$
    and containing the box $s$, is moved under the column in $F(b)$
    lying at the position $l_{h}(s) \tmodulus{(p_{h})}$. Thus, in both cases,
    the column in $F(b)$ at the position $l_{h}(s)\tmodulus{(p_h)}$
    contains the box $s$. On the other hand, the number of moved columns
    depends on whether $l_{h}(s) \tmodulus{(p_{h})}$ is smaller than
    $(b-b_{h}) \tmodulus{(p_{h})}$ or not.
    If $l_{h}(s) \tmodulus{(p_{h})} < (b-b_{h}) \tmodulus{(p_{h})}$ then the number
    of columns of length $T^{(2)}_{h}$, moved under the column at
    the position $l_{h}(s)\tmodulus{(p_h)}$, is $(\lfloor\frac{b-b_{h}-1}{p_{h}}\rfloor)$.
    If $l_{h}(s) \tmodulus{(p_{h})} \geq (b-b_{h}) \tmodulus{(p_{h})}$ then one needs
    to move one more column of length $T^{(2)}_{h}$, so their total
    number is $(\lfloor\frac{b-b_{h}-1}{p_{h}}\rfloor +1)$.
\end{description}

\noindent From the above considerations, we finally infer the following%

\begin{theorem}
Assume $0<\mu<1$ and $\mu\in\mathbb{Q}$. Let the roots of the
algebraic equation (\ref{algebraic curve 0<p<q}) be labeled as
described in Subsection \ref{Sec:mu-0<mu<1-rational:Topological}.
Let $T(s,b)$ be the period of $w_s$, the $s$-th root of the
algebraic equation (\ref{algebraic curve 0<p<q}) when $\xi$ moves
along a closed path on the $\xi$-plane, including $b$ consecutive
adjacent branch points starting from the branch point
$\xi^{(1)}_{\mathrm{b}}$. Let $h$ be the integer such that $0\leq
b_{h} \leq b < b_{h+1}\leq q-1$ and $l_{h}(s)$ the symbol
(\ref{lsymbol}). Let $T^{(1)}_{h}$ and $T^{(2)}_{h}$ be the
quantities described by the recursions (\ref{T1 T2}). Then we have
the following period formula, for all $b<q$:
\begin{subequations}
\begin{equation} \label{Period Formula 0<mu<1 mu rational-a}
T(s,b)=\left\{
         \begin{array}{ll}
         T^{(1)}_{h} + \left ( \left \lfloor \frac{b-b_{h}-1}{p_{h}} \right \rfloor \right
         ) \, T^{(2)}_{h} & \mbox{ if } l_{h}(s)\leq b-b_{h}+p_{h} \mbox{ and } \\
         & (b-b_{h}) \tmodulus{(p_{h})} < l_{h}(s) \tmodulus{(p_{h})}\\
         & \\
         T^{(1)}_{h} + \left ( \left \lfloor \frac{b-b_{h}-1}{p_{h}} \right \rfloor +1 \right
         ) \, T^{(2)}_{h} & \mbox{ if } l_{h}(s)\leq b-b_{h}+p_{h} \mbox{ and } \\
         & (b-b_{h}) \tmodulus{(p_{h})} \geq l_{h}(s) \tmodulus{(p_{h})}\\
         & \\
         T^{(2)}_{h} & \mbox{ if } l_{h}(s) > b-b_{h}+p_{h} \mbox{ .}\\
         \end{array}
         \right.
\end{equation}
\noindent If $b=q$ we have
\begin{equation} \label{Period Formula 0<mu<1 mu rational-b}
  T(s,q)=\left\{
         \begin{array}{ll}
         q-p & \mbox{ if } 1\leq s \leq q-p\\
         &\\
         p & \mbox{ if } q-p+1\leq s \leq q \mbox{ .}\\
         \end{array}
         \right.
\end{equation}
\end{subequations}
\noindent Via the connection with continued fractions, as shown in
Subsection \ref{Sec:mu-0<mu<1-rational:continued fractions}, we can
reformulate (\ref{Period Formula 0<mu<1 mu rational-a}) as follows:
\begin{equation} \label{Period Formula with Continued Fractions 0<mu<1 mu rational}
T(s,b)=\left\{
         \begin{array}{ll}
         P_{h-2} + \left ( \left \lfloor \frac{b-b_{h}-1}{q-b_{h+1}} \right \rfloor +1 \right
         ) \, P_{h-1} & \mbox{ if } l_{h}(s)\leq b+q-(b_{h}+b_{h+1}) \mbox{ and }\\
         {}& (b-b_{h}) \tmodulus{(q-b_{h+1})} < l_{h}(s) \tmodulus{(q-b_{h+1})}\\
         {}&{}\\
         P_{h-2} + \left ( \left \lfloor \frac{b-b_{h}-1}{q-b_{h+1}} \right \rfloor +2 \right
         ) \, P_{h-1} & \mbox{ if } l_{h}(s)\leq b+q-(b_{h}+b_{h+1}) \mbox{ and }\\
         {}& (b-b_{h}) \tmodulus{(q-b_{h+1})} \geq l_{h}(s) \tmodulus{(q-b_{h+1})}\\
         {}&{}\\
         P_{h-1} & \mbox{ if } l_{h}(s) > b+q-(b_{h}+b_{h+1}) \mbox{ ,}\\
         \end{array}
         \right.
\end{equation}
\noindent where $P_k$ is the numerator of the $k$-th convergent of
the continued fraction expansion of $\frac{1}{1-\mu}$.
\end{theorem}

\noindent For any inclusion of consecutive branch points, the final
formula (\ref{Period Formula with Continued Fractions 0<mu<1 mu
rational}) indicates that the roots of the algebraic equation
(\ref{algebraic curve 0<p<q}) can feature only three possible
periods (see Table 1). Note that the sum of
the first and third expressions in the right hand side of this
formula, (\ref{Period Formula with Continued Fractions 0<mu<1 mu
rational}), always gives the second expression. Moreover, if $b$
takes the special values $b=b_{h}+n\,(q-b_{h+1}) \,\,,\,\, 0\leq n
\leq a_{h}-1 \,\,,\,\, n\in\mathbb{N}$, where $a_h$ is the $h$-th
partial quotient appearing in the continued fraction expansion of
$\frac{1}{1-\mu}$, then the roots of (\ref{algebraic curve 0<p<q})
can feature only two possible periods: indeed, in these particular
cases, the first condition in (\ref{Period Formula with Continued
Fractions 0<mu<1 mu rational}) fails for all $s$ since $(b-b_{h})
\tmodulus{(q-b_{h+1})}=q-b_{h+1}$ and $1 \leq
l_{h}(s)\tmodulus{(q-b_{h+1})} \leq q-b_{h+1}$.

Summarizing, given the initial data $z_i(0), i=1,2,3$, and the
coupling constants $f,g$ (such that $\mu\in\mathbb{Q}$) we
can now predict the period of the function $\check w(t)$ by the
following prescription:
\begin{enumerate}
\item Draw the circles $\Xi$ and $B$ defined by \eqref{LocMovableSing-ksi} and \eqref{ksi}.
\item Identify the label of the physical root $\check s$ following the prescription given in Section \ref{Sec:mu-0<mu<1-rational:Physical Root}.
\item Calculate the number of branch points $b$ included in the circle $\Xi$.
\item Develop in continued fraction the ratio $\frac{1}{1-\mu}$ and build the sequence $b_k$ defined by \eqref{b sequence in terms of P and Q}. Identify the element $h$ of the sequence such that $b_h\leq b<b_{h+1}$.  Calculate $l_h(\check s)$ using \eqref{lsymbol}.
\item The period  $T(\check s, b)$ of $\check w(t)$ is given by Theorem 1.
\end{enumerate}

The understanding of the topology of the Riemann surface $\Gamma$ for
$\mu\in\mathbb{Q}$ allows one to associate the evolution in time of
the physical problem \eqref{EqMot} to the symbolic dynamics given by the
sequence of the labels of the visited sheets of $\Gamma$. In other words, one can describe
the \textit{time-$T$ map} obtained by sampling the solution at every time interval $T$, as a sequence of natural
numbers, \textit{i.e.} the sequence of numbers labeling the visited
sheets, after each round trip as $\xi$ travels on the Riemann
surface.

\begin{center}
\begin{table}\label{Tab:T(s,b) p=5 q=12}
\begin{footnotesize}
$$
\begin{array}{||c||c|c|c|c|c|c|c|c|c|c|c|c||}
  \hline
  \hline
  s = & 1 & 2 & 3 & 4 & 5 & 6 & 7 & 8 & 9 & 10 & 11 & 12\\
  \hline
  \hline
  b=0 & 1 & 1 & 1 & 1 & 1 & 1 & 1 & 1 & 1 & 1 & 1 & 1 \\
  b=1 & 2 & 1 & 1 & 1 & 1 & 1 & 1 & 2 & 1 & 1 & 1 & 1 \\
  b=2 & 2 & 2 & 1 & 1 & 1 & 1 & 1 & 2 & 2 & 1 & 1 & 1 \\
  b=3 & 2 & 2 & 2 & 1 & 1 & 1 & 1 & 2 & 2 & 2 & 1 & 1 \\
  b=4 & 2 & 2 & 2 & 2 & 1 & 1 & 1 & 2 & 2 & 2 & 2 & 1 \\
  b=5 & 2 & 2 & 2 & 2 & 2 & 1 & 1 & 2 & 2 & 2 & 2 & 2 \\
  b=6 & 3 & 2 & 2 & 2 & 2 & 3 & 1 & 3 & 2 & 2 & 2 & 2 \\
  b=7 & 3 & 3 & 2 & 2 & 2 & 3 & 3 & 3 & 3 & 2 & 2 & 2 \\
  b=8 & 5 & 3 & 5 & 2 & 2 & 5 & 3 & 5 & 3 & 5 & 2 & 2 \\
  b=9 & 5 & 5 & 5 & 5 & 2 & 5 & 5 & 5 & 5 & 5 & 5 & 2 \\
  b=10 & 7 & 5 & 7 & 5 & 7 & 7 & 5 & 7 & 5 & 7 & 5 & 7 \\
  b=11 & 12 & 12 & 12 & 12 & 12 & 12 & 12 & 12 & 12 & 12 & 12 & 12 \\
  b=12 & 7 & 7 & 7 & 7 & 7 & 7 & 7 & 5 & 5 & 5 & 5 & 5 \\
  \hline
  \hline
\end{array}
$$
\end{footnotesize}
\caption{Values of $T(s,b)$ when $p=5$ and $q=12$.}
\end{table}
\end{center}

\subsection{The period formula for $\mu\notin\mathbb{Q}$}\label{Sec:mu-0<mu<1-irrational}%

We treat the case in which $\mu$ is an irrational number as
a limit of the case in which $\mu$ is a rational number. If
$\mu=p/q$ gets close to being irrational, the number $q$ -- that
corresponds to the number of roots of equation (\ref{algebraic curve
0<p<q}) as well as the number of branch points of $w(\xi)$ in the
complex $\xi$-plane -- approaches infinity. So we must analyze all
the previous formulae in the limit:
\begin{equation}\label{limit 0<mu<1}
    p,q\rightarrow\infty \,\, \mbox{ with } \,\, 0<\mu=\frac{p}{q}<1 \,,
\end{equation}
\noindent namely in the limit in which the integers $p$ and $q$
diverge so that their ratio tends to the irrational number $\mu$.

If $\mu$ is an irrational number, it follows that the Riemann
surface $\Gamma$ associated to (\ref{Eqw}) becomes an
$\infty$-sheeted covering of the complex $\xi$-plane. In the limit
(\ref{limit 0<mu<1}), the continued fraction expansion of
$\frac{1}{1-\mu}$ has length $\bar{k}$ that tends to infinity:
$\bar{k}\rightarrow\infty$. The first step is to normalize
the $q$-sequence (\ref{q sequence}), introducing the new sequence
$\{\rho_{k}\}\sim\left\{\frac{q_{k}}{q}\right\}$:
\begin{equation}\label{rho sequence}
\rho_{k}=\frac{\rho_{k-1}}{r_{k-1}} \,,\qquad \rho_{0}=1 \,,\qquad
\rho_{1}=1-\mu \,,
\end{equation}
\noindent where $r_{k}$ is the sequence of the (irrational)
remainders of the continued fraction of $\frac{1}{1-\mu}$ (see
(\ref{r sequence})). In the limit (\ref{limit 0<mu<1}), the
$\rho$-sequence becomes a strictly decreasing sequence of irrational
numbers such that $\rho_{k}\rightarrow 0$ as $k\rightarrow\infty$.

\begin{figure}
\centering
\includegraphics[width=5cm]{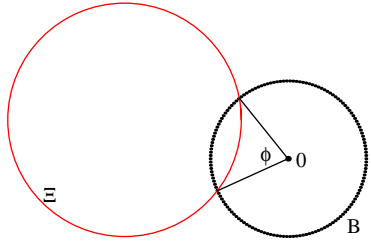}
\caption{The definition of the angle $\phi$}\label{Fig:AngoloPhi}
\end{figure}

The next step is to obtain the proper continuous variable which
replaces the discrete index $b$. Let $B$ be the circle on which the
square-root branch points lie and $\Xi$ the circle on which the
$\xi$ variable travels periodically, see (\ref{ksi}). We define the
variable $\nu$, which is $0$ if $\Xi$ does not contain any arc of
$B$ and $1$ if $\Xi$ does contain the whole circle $B$. If $B$ and
$\Xi$ intersect, select the arc of the $B$ circle which is contained
inside the evolutionary circle $\Xi$ and denote by $\nu$ the
corresponding angle $\phi$, normalized by $2\pi$ (see Figure
\ref{Fig:AngoloPhi}). Since
\begin{equation}\label{nu}
    \nu=\frac{\phi}{2\pi}=\frac{b}{q} \,,
\end{equation}
\noindent this is a convenient continuous variable appropriate to
replace $b$ in the case $\mu$ is an irrational number.

We then normalize the fundamental $b$-sequence (\ref{b sequence in
terms of P and Q}), replacing it by the new sequence
$\{\nu_{k}\}\sim\left\{\frac{b_{k}}{q}\right\}$: \begin{subequations}
\begin{equation}\label{nu sequence}
\nu_{k}=1-\rho_{k} \,,\qquad \nu_{0}=0 \,,\quad k\in\mathbb{N}\,;
\end{equation}
\noindent in the limit (\ref{limit 0<mu<1}), we have that
$\nu_{k}\rightarrow 1$ as $k\rightarrow\infty$. It is also
convenient to reformulate $\nu_{k}$ using the language of continued
fractions. From (\ref{b sequence in terms of P and Q}), in the limit
(\ref{limit 0<mu<1}) we have:
\begin{equation}\label{nu sequence in terms of P and Q}
\nu_{k}=1-(-1)^{k}\left[(\mu-1)\,P_{k-2}+Q_{k-2}\right] \,,\qquad
k\in\mathbb{N} \,.
\end{equation}
\end{subequations}

We are finally ready to enunciate the period formula when $\mu$ is
an irrational number.

\begin{theorem}
Let $0<\mu<1$ and $\mu\notin\mathbb{Q}$. Let $T(\nu)$ be the period
of one of the roots of the algebraic equation (\ref{Eqw}) when $\xi$
travels on the circle $\Xi$ in the complex $\xi$-plane, intersecting
the square-root branch points circle $B$ in such a way that
$0\leq\nu<1$ (see (\ref{nu})). Let $h$ be the integer such that
$0\leq \nu_{h} \leq \nu < \nu_{h+1}<1$. Let $P_{k}$ be the numerator
of the $k$-th convergent of the continued fraction expansion of
$\frac{1}{1-\mu}$ (see (\ref{convergent numerator and
denominator})). If $\nu=0$, then $T(\nu)=1$. Otherwise, $T(\nu)$
takes one of the following three values:
\begin{equation} \label{Period Formula with Continued Fractions 0<mu<1 mu irrational}
  T(\nu)=\left \{
         \begin{array}{l}
         P_{h-2} + \left ( \left \lfloor \frac{\nu-\nu_{h}}{1-\nu_{h+1}} \right \rfloor +1 \right
         ) \, P_{h-1},, \\
         \\
         P_{h-2} + \left ( \left \lfloor \frac{\nu-\nu_{h}}{1-\nu_{h+1}} \right \rfloor +2 \right
         ) \, P_{h-1} \,,\\
         \\
         P_{h-1} \,.\\
         \end{array}
         \right.
\end{equation}
\noindent If $\nu=1$, namely if the evolutionary circle $\Xi$
contains entirely the square-root branch points circle $B$, then the
dynamics becomes again simple, because the evolutionary curve
effectively surrounds only the irrational branch point at infinity.
In this latter case, the time evolution of the generic root of
(\ref{Eqw}) is quasi-periodic, involving a (nonlinear)
superposition of two periodic evolutions with noncongruent periods,
$1$ and $\frac{1}{\mu}$ or $1$ and $\frac{1}{1-\mu}$, as can be
easily understood observing the exponents of the asymptotics
(\ref{Ansatz1a})-(\ref{Ansatz1c}) and
(\ref{Ansatz2a})-(\ref{Ansatz2c}).
\end{theorem}

\noindent Note that, as in the rational case, the sum of the first
and third expressions in the right hand side of (\ref{Period Formula
with Continued Fractions 0<mu<1 mu irrational}) always gives the
second expression. We remark that in the irrational case there are
no labels for the roots, so we cannot assign the periods to the
roots in terms of their labels as we did in the rational case.
Therefore, a specific prescription to identify the period of the
physical root $\check w(t)$ cannot be given. At best, we can assure
that the period is one of the three possibilities specified by
(\ref{Period Formula with Continued Fractions 0<mu<1 mu
irrational}).

For $\nu=0$, the period $T(0)$ is equal to $1$. Since $1$ is the
accumulation point of the sequence $\{\nu_{k}\}$, if $\nu$ is
instead close to $1$ (\emph{i.e.} if the evolutionary circle $\Xi$
contains almost completely the branch points circle $B$), a small
change in the time trajectory results in a drastic change of the
observed periods. In fact, $T(\nu)$ approaches infinity as
$\nu\rightarrow 1$ and the values of $T(\nu)$ depend on the partial
quotients $a_{k}$ of the continued fraction expansion of
$\frac{1}{1-\mu}$; such partial quotients are well-known to be
chaotic and unpredictable in their sequence for a generic irrational
number (for almost all the irrational numbers, except for the
quadratic irrationals that feature periodic continued fraction
expansions).

In the next subsection we treat a remarkable example where it is possible to obtain explicitly the
asymptotic behavior of $T(\nu)$ as $\nu\sim 1$.

\subsection{A remarkable example}\label{Sec:mu-0<mu<1-irrational:Remarkable example}%

In this subsection we display the special example with the following
(conveniently chosen) quadratic irrational value of $\mu$
in the interval $0<\mu<1$:
\begin{subequations}
\begin{equation}\label{goldenratio special mu 0<mu<1}
    \mu=\frac{2}{3+\sqrt{5}}=\frac{1}{1+\varphi} \,,
\end{equation}
\noindent such that
\begin{equation}\label{goldenratio}
    \frac{1}{1-\mu}=\varphi=\frac{1+\sqrt{5}}{2} \,,
\end{equation}
\noindent where $\varphi$ is the so-called \textit{golden ratio},
namely the positive solution of the second degree equation
\begin{equation}\label{goldenratio equation}
   \varphi^{2}-\varphi-1=0 \,.
\end{equation}
\end{subequations}

The golden ratio $\varphi$ has the nice property that all the
infinite partial quotients (\ref{a sequence}) appearing in its
continued fraction expansion are equal to unity:
\begin{equation}\label{goldenratio as continuedfraction}
\varphi=1+\frac{1}{1+\frac{1}{1+\frac{1}{1+\frac{1}{...}}}}
\,,\qquad a_{k}=1 \,\, \forall\,k \geq 0 \,.
\end{equation}
\noindent So, in this particular case, the recursion relation
(\ref{r sequence}) becomes:
\begin{equation}\label{goldenratio r sequence}
r_{k}=\frac{1}{r_{k-1}-1} \,\,\mbox{ with }\,\, r_{0}=\varphi
\,\,,\,\, k\geq 1 \,.
\end{equation}
\noindent By combining this last relation with (\ref{rho sequence})
we get:
\begin{equation}\label{goldenratio rho sequence}
\rho_{k+1}=\rho_{k-1}-\rho_{k} \,\,\mbox{ with }\,\, \rho_{0}=1
\,\,,\,\, \rho_{1}=\frac{1}{\varphi}=\varphi-1 \,\,,\,\, k\geq 1
\end{equation}
\noindent and from this, via (\ref{nu sequence}),
\begin{equation}\label{goldenratio nu sequence}
\nu_{k+1}=\nu_{k-1}-\nu_{k}+1 \,\,\mbox{ with }\,\, \nu_{0}=0
\,\,,\,\, \nu_{1}=1-\frac{1}{\varphi} \,\,,\,\, k\geq 1 \,.
\end{equation}
\noindent Solving this last relation with respect to $k$, we get
\begin{equation}\label{goldenratio nu solution}
\nu_{k}=1-\varphi^{-k} \,\,,\,\,k \geq 0 \,.
\end{equation}

Moreover -- somewhat remarkably -- (\ref{goldenratio as
continuedfraction}) implies, via (\ref{convergent numerator and
denominator}), the following relation for the numerators of the
convergents:
\begin{equation}\label{goldenratio numerator convergents}
P_{k}=P_{k-1}+P_{k-2} \,\,\,,\,\,\, P_{-2}=0 \,\,\,,\,\,\, P_{-1}=1
\,\,\,,\,\,\, k \geq 0 \,,
\end{equation}
\noindent  \emph{i.e.} exactly the recurrence relation for
Fibonacci's numbers. Using Binet formula, we have:
\begin{equation}\label{goldenratio Fibonacci}
P_{k}=\frac{1}{\sqrt{5}}\left[\varphi^{k+2}-(-\varphi)^{-(k+2)}\right]
\,\,\,,\,\,\, k \geq -2 \mbox{ .}
\end{equation}

Through (\ref{goldenratio nu solution}), we can explicitly invert
the inequality $\nu_{k}\leq\nu<\nu_{k+1}$, finding, for a fixed
value of $\nu$ in the interval $0<\nu<1$, the integer number $k$
such that $\nu_{k}\leq\nu<\nu_{k+1}$:
\begin{equation}\label{goldenratio kappa}
k\equiv k(\nu)=\left\{
\begin{array}{ll}
- \left \lfloor \frac{\log(1-\nu)}{\log(\varphi)} \right \rfloor -1\,\,\,, & \mbox{if } 0<\nu<1 \mbox{ ;} \\
\\
0\,\,\,, & \mbox{if } \nu=0 \,. \\
\end{array}
\right.
\end{equation}
\noindent Using (\ref{goldenratio kappa}), we see that the argument
of the floor function in (\ref{Period Formula with Continued
Fractions 0<mu<1 mu irrational}) must satisfy the inequalities:
\begin{equation}\label{goldenratio disuguaglianza}
0<\frac{\nu-\nu_{k(\nu)}}{1-\nu_{k(\nu)+1}}<\varphi-1<1 \,\,\,\mbox{
;}
\end{equation}
\noindent so the floor function in (\ref{Period Formula with
Continued Fractions 0<mu<1 mu irrational}) always vanishes and the
root period $T(\nu)$ has one of the following values:
\begin{equation}\label{goldenratio period values}
T(\nu)=\left\{P_{k+1},P_{k},P_{k-1}\right\} \,\,\,\mbox{ with
}\,\,\, 0\leq\nu_{k}\leq\nu<\nu_{k+1}<1 \,,
\end{equation}
\noindent namely one of three consecutive Fibonacci's numbers. From
(\ref{goldenratio numerator convergents}), via (\ref{goldenratio
kappa}), we see that, for $0<\nu<1$,
\begin{equation}\label{goldenratio diseguaglianze P}
\frac{1}{\sqrt{5}}\left[ \frac{\varphi}{1-\nu}-\frac{1-\nu}{\varphi}
\right] < P_{k(\nu)} < \frac{1}{\sqrt{5}}\left[
\frac{\varphi}{1-\nu}+\frac{1-\nu}{\varphi} \right]\,.
\end{equation}

From the above relations (\ref{goldenratio diseguaglianze P}) and
(\ref{goldenratio disuguaglianza}), and from the period formula
(\ref{Period Formula with Continued Fractions 0<mu<1 mu
irrational}), we obtain the following lower and upper bounds for the
period values in terms of $\nu$ in the interval $0<\nu<1$:
\begin{equation}\label{goldenratio lower and upper bound of the period}
\frac{\nu\,(2-\nu)}{\sqrt{5}\,(1-\nu)}\leq
T(\nu)<\frac{7+\sqrt{5}+(\sqrt{5}-3)\,\nu\,(2-\nu)}{2\,\sqrt{5}\,(1-\nu)}\,.
\end{equation}
\noindent These inequalities entail that the integer $T(\nu)$
diverges proportionally to $(1-\nu)^{-1}$ as $\nu\rightarrow 1$.

\section{Summary and conclusions}

We have studied the trajectories of the system of three coupled ODEs
\eqref{EqMot} in the semi-symmetrical case \eqref{Symm} when two
coupling constants are equal, and the ratio $\mu$ defined in
\eqref{mu} belongs to the interval $(0,1)$.

For rational values of $\mu$, all orbits are periodic for arbitrary
initial data, except for a set of null measure that corresponds to a
collision in finite time. Moreover, the system is
\textit{isochronous}: the periodic orbits are neutrally stable and a
small perturbation of the initial condition produces another
periodic orbit of the same period. In this case, the Cauchy
problem can be solved explicitly and the evolution in time of the
physical trajectories can be put in correspondence with the symbolic
dynamics (a sequence of natural numbers) of the labels of the
visited sheets of the associated Riemann surface.

For irrational values of the ratio $\mu$, one must distinguish two
cases depending on the relative position of the evolutionary circle
$\Xi$, defined by \eqref{ksi}, with respect to the branch-point
circle $B$, defined by \eqref{LocMovableSing-ksi}. If $B$ is
entirely contained in $\Xi$, then the solution is quasi-periodic,
while if $\Xi$ contains only a part of $B$, then almost all orbits
are periodic (again, except for a set of null measure that
corresponds to a collision in finite time). The relative position of
these two circles depends on the choice of initial data. In fact,
initial data can always be found such that the period of the
corresponding orbit is as high as desired (although always a
multiple of the fundamental period). This situation corresponds to
the limiting case between the two cases described above. An explicit
instance of this situation has been given for a particularly simple
example related to the golden ratio and Fibonacci's numbers.

The main result of this paper is to derive explicit formulae for the
period as a function of the initial data and the coupling constants.
The only way the period of an orbit can change as the initial data
are varied is passing through a singularity (corresponding to a
collision between the particles). The dependence of the period on
the initial data is shown to depend on the continued fraction
expansion of the parameter $\mu$, which for irrational values shows
a rich  behavior. We would like to emphasize the novelty of the
approach developed in this paper: it is not possible to achieve such
an explicit description of the dynamics of the system using the
qualitative theory of ODEs, which illustrates why the complete
description of the Riemann surface is an important goal in the cases
where it can be achieved. The description performed in this case
involves the use of techniques that range from complex analysis and
geometry to graph theory and continued fractions, which shows the
novelty of the approach.

One of the important lessons from the results in this paper in relation to the ideas of Kruskal, Bountis, Tabor and their collaborators mentioned in the Introduction is the following: the fact that the Riemann surface associated to the solution of the problem is infinitely sheeted is not enough for a system to produce chaotic behaviour, although it can be a source of unpredictable long term behaviour. This is illustrated by the behaviour of the semi-symmetric three-body problem with an irrational value of $\mu$ (i.e. generic values of the coupling constants). In this case, we have shown that the Riemann surface is infinitely-sheeted and it contains an infinite number of branch points whose position depends on the initial data and whose projection on the complex plane fills densely a circle $B$. Furthermore,we have proved that all solutions are either periodic, quasi-periodic or lead to a collision in finite time. Despite this fact, even in the periodic case there is a source of unpredictability associated to the fact that it is not possible to identify the root associated to the physical problem among the infinity of roots of \eqref{Eqw}. Moreover, two trajectories originating from nearby initial data will separate at some future time as they will travel to different sheets on the Riemann surface, corresponding to the fact that the \textit{collision manifolds} (locus of initial data in phase space that lead to a collision in finite time) fill densely an open set of phase space.
%
On the other hand, in the $\mu$-rational case, not only can the period be predicted by an explicit formula, but the complete motion can be described by the sequence of sheets in the Riemann surface that are visited during the motion.

The description of the remaining case $\mu>1$ can be tackled with similar techniques to the ones developed in this paper. However, we have decided to postpone this analysis to a further publication, where the sensitive dependence on the initial conditions will be discussed. An interesting further step in this program would be to understand how sensitive dependence or classical indicators of chaos such as Lyapunov exponents are related to properties of the Riemann surface.

\vskip 0.6cm
\noindent \textbf{Acknowledgements.} It is a pleasure to acknowledge illuminating discussions
with Carl Bender, Boris Dubrovin, Yuri Fedorov, Jean-Pierre
Fran\c{c}oise, Peter Grinevich and  Fran\c{c}ois Leyvraz.
The research of DGU was supported in part by MICINN-FEDER grant
MTM2009-06973 and CUR-DIUE grant 2009SGR859 and he would like to
thank the financial support received from the Universit\`a di Roma
``La Sapienza'' under the Accordo Bilaterale with Universidad
Complutense de Madrid.

\end{document}